\documentclass[a4paper,11pt,amsmath,mathrsfs,placeins]{article}
\usepackage{jcappub} 

 \newcommand{\beq}[1]{\begin{equation}\label{#1}}
 \newcommand{\eeq}{\end{equation}}
 \newcommand{\bea}[1]{\begin{eqnarray}\label{#1}}
 \newcommand{\eea}{\end{eqnarray}}

\title{\boldmath Linear Stability Analysis of Evolving Thin Shell Wormholes}
 \author{Ai-chen Li,}
 \author{Wu-long Xu~\&}
 \author{Ding-fang Zeng \footnote{Corresponding author.}}
 \affiliation{Theoretical Physics Division, Beijing University of Technology, Beijing 100124}
\emailAdd{lac@emails.bjut.edu.cn}
\emailAdd{wlxu@emails.bjut.edu.cn}
\emailAdd{dfzeng@bjut.edu.cn}

\abstract{Using ideas of cosmological perturbation theories, we make general linear stability analysis of dynamic thin-shell wormholes constructed by cutting-and-pasting two building-block spacetimes at arbitrary joining shell radius. We observed that in appropriate parameter choices, dynamical thin shell wormholes following from such a cut-and-paste procedure can be kept stable during the whole evolution process towards the final extremal point on which the joining shell radius reaches on static values. Our work forms a valuable complementarity to previous analysis basing on virtual spherical perturbations around the born-static configuration of the wormholes.}

\begin{document} 
\maketitle
\flushbottom

\section{Introduction}
Wormhole is a special structure bridging two asymptotic regions in one universe or two universes in the multiverse. To implement a traversable wormhole \cite{Morris:1988cz, HGEllis1973, Hochberg:1997wp, Visser1995}, the averaged null energy condition (ANEC) has to be violated \cite{Morris:1988tu,Hochberg:1998ii,Hochberg:1998vm}. In the old Morris-Thorne type wormholes \cite{Morris:1988cz,HGEllis1973}, ``exotic'' i.e. ANEC-violating matters are distributed in the whole spacetime and the spacetime are required to be asymptotically flat and spheric symmetric. While in the Visser type thin-shell wormholes (TSW)  \cite{Visser:1989kh,Visser:1989kg}, a so called cut-and-paste method is introduced and the necessity for energy condition violation and asymptotic feature of the spacetimes are heavily reduced \cite{Hochberg:1997wp}. For example, exotic matters in them are filled only in the thin-shell joining the two building-block spacetimes, asymptotics of the building-blocks are no more limited to static and spherical symmetric \cite{Bejarano:2006uj,Bhawal:1992sz,Hochberg:1990is,Agnese:1995kd, Eiroa:2005pc,Lemos:2004vs,Clement:1997yp,Lobo:2005us,Lobo:2005vc}. In these works, evolutions of the joining shell radius are determined by a Friedman like equation $\dot{r}^2\equiv H[r,\cdots]$, ellipsis here denoting parameters related with the feature of exotic filling matter and building-block spacetime asymptotics.

Obviously, a wormhole's stability and lifetime have key senses for practical spacetime traveling. By spherical symmetric perturbation analysis \cite{Poisson:1995sv} of simple TSWs, Visser observed that such question is directly related with the equation of state of matters filled in the joining shell. While in \cite{Lobo:2003xd}, it is found that the asymptotics of the building-block spacetimes also have effects on the stability of perturbations. Then in \cite{Lemos:2008aj, Sharif:2013efa,Garattini:2008xz,Lobo:2009ip,Garcia:2010xb}, more analysis are made for various building-block spacetimes with results consistent with \cite{Poisson:1995sv,Lobo:2003xd}.  While \cite{Lobo:2005yv,Eiroa:2007qz,Gorini:2008zj,Halilsoy:2013iza,Harko:2014oua} make analysis for various filling matters' equation of state and figure out stability regions in the appropriate parameter space. However, in all these works, stability is defined relative to a born-static point satisfying $H(r_0)=0$, $H'(r_0)=0$, $H''(r_0)<0$, around which the joining shell could not experience true evolutions at all.  Obviously, a more natural question arises here, if the TSW is cut-and-paste at a non-born-static point of the joining-shell radius, will it experience a dynamical evolution and arrive on a final stable point defined by $H=0$, $H'=0$, $0<H''$? Additionally, if a cut-and-paste procedure is executed in such an environment that no static point exists, then how to judge the resultant TSW is stable or not? What's more, quantum fluctuations occur everywhere, we have no reason to exclude matter fluctuations occurring on the joining-shell, which will induce perturbations to the wormhole unavoidably. The purpose of this work is to provide necessary formulas for stability analysis of thin-shell wormholes in dynamic evolution. Our basic strategy is borrowing ideas from the cosmological perturbation theory, especially brane world cosmologies \cite{Garriga:1991ts,Brax:2001qd, Guven:1993ew, Ishibashi:2002nn, Boehm:2002kf, Deruelle:2000yj,Brax:2002nx}. Our results indicate that, at least to linear order in perturbations, a TSW cut-and-paste on a non-born-static point of the joining-shell radius can experience a natural and robust  evolution to reach a stable final configuration in appropriate parameter choices.

The building-block spacetimes of this paper are organized as follows. Section \ref{MEofTSW} is necessary introductions to cut-and-paste methods for TSW's construction and the resultant wormholes' dynamic evolution. We will consider two kinds of filling matters, i.e. Phantom Gas and Chaplygin Gas in the joining shell and three kinds of building-block spacetime, Schwarzchild-Anti-deSitter (SAdS), Schwarzchild-deSitter (SdS) and Hayward blackhole with regular center \cite{Hayward:2005gi} respectively. Section \ref{PerofETSW} is our perturbation formulas developing and stability criteria discussion for TSWs with dynamically evolving joining shell. We will focus on scalar type perturbations but will not constrain to spherical symmetric modes. Section \ref{Nresults} contains our numerical results for stability analysis in some typical parameter choices. The final section is our conclusion and prospects for future works.

\section{Thin shell wormholes and their evolution}
\label{MEofTSW} 

We will use the now standard cut-and-paste method \cite{Visser:1989kh,Visser:1989kg} to construct the so called thin shell wormholes. As the first step, we take two building-block spacetimes each containing a static blackhole, remove from each of them an n-dimensional (we will focus on $n=4$) sub-region 
\bea{}
\Sigma_{1,2}=\{\mathrm{sub}M_{1,2}|r\leqslant r_{cut}~\mathrm{with}~r_h<r_{cut}\}
\eea
where $r_h$ is the horizon radius of the blackhole, and $r_{cut}$ is the radius of the removed regions. This two region has boundaries defined by $\partial \Sigma_{1,2}$,
\bea{}
\partial \Sigma_{1,2}=\{\mathrm{sub}\Sigma_{1,2}|r= r_{cut}~\mathrm{with}~r_h<r_{cut}\}
\eea
which are two (n-1)-dimensional timelike hypersurfaces. Secondly, we join $\Sigma_1$ and $\Sigma_2$ together with a thin shell filled with matters defined by their equation of state, at the same time identify the two (n-1)-dimensional timelike hypersurfaces (namely $\partial\Sigma=\partial \Sigma_1=\partial \Sigma_2$). After this two steps, we get a manifold $\mathcal{M}$ possessing two asymptotic regions connected by a wormhole, with the throat denoted by symbol $\partial \Sigma$. Due to the $Z_2$ symmetry between $\Sigma_{1,2}$, we will denote $\Sigma_{1,2}$ with $\Sigma_{\pm}$ in the following. 

\begin{figure}[!ht]
\begin{center}
\includegraphics[scale=0.8]{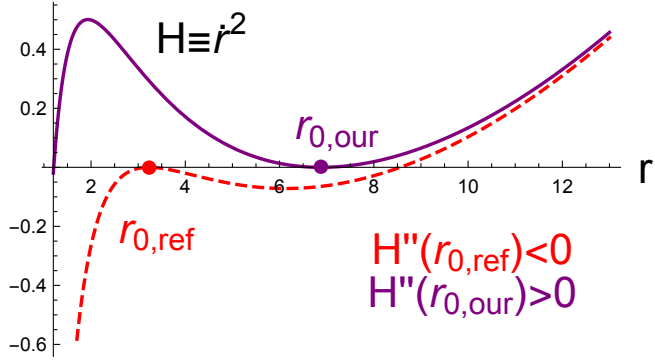}
\caption{(color online) The previous works use conditions $H(r_0)=0$, $H'(r_0)=0$, $H''(r_0)<0$ to define and look for stable configurations in the cut-and-paste method. The resulting wormhole is born-static and stable. While our goal in this work is to find configurations so that $H(r_0)=0$, $H'(r_0)=0$, $0<H''(r_0)$. The cut-and-paste is executed on non-$r_0$ value, but along the evolution way, the wormhole keeps stable to general inhomogeneous and anisotropic perturbations.}
\label{figEPComRef}
\end{center}
\end{figure} 

Obviously, the wormhole constructed this way is a time-like thin shell $\partial \Sigma$. In general cases, this shell cannot be static. To get static configuration, the common doing of previous works in this area  is to choose/fine-tuning either theoretic frameworks of the building-block spacetime or the equation of state of matters filling in the joining shell. However, as long as we know, all previous works focus on parameter combinations so that the resulting wormhole are born-static, which are defined by conditions $\dot{r}^2(\equiv H)=0$, $H'(r)=0$ and $H''(r)<0$ in the $H-r$ plane. While our goal in this work is to find wormholes connected by joining shells which are dynamic and evolving with time, but are stable to general inhomogeneous perturbations, see figure \ref{figEPComRef} for references. These should be more natural wormholes. So unlike those in previous works, our stability analysis in this work is an analysis similar to cosmological perturbations occurring on an evolving thin shell.

\subsection{Motion of thin shells in a static bulk with one extra dimension \label{TSinBulk}}

To accomplish our goal, we need first know about the thin shell themselves' evolution. We introduce coordinating system so that the bulk space-time is denoted by $X^M$ while the shell, by $x^\mu$, see Table.\ref{EmbedBrane} for references. Evolutions of the shell in the bulk space-time will sweep out a world-volume characterized by $X^M(x^\mu)$, we can define the vielbein $e^\mu_M =(\frac{\partial X^M}{\partial x^\mu})^{-1}$ as the perspective  translation tool from the shell observer to the bulk ones, and vice versa. 
\begin{table}[!ht]
\begin{center}
\begin{tabular}{|c|c|c|c|}
\hline
Bulk & $T$ & $r$ & $\vec{X}_2$ \\
\hline
thin shell & $t$ & $\times$ & $\vec{x}_2$ \\
\hline
embedding  & $T(t)$ & $r(t)$ & $\vec{x}_2=\vec{X}_2$\\
\hline
\end{tabular}
\caption{Coordinates parameterizing the whole bulk spacetime and the thin shell in it.}
\label{EmbedBrane}
\end{center}
\end{table}
Under the so called static gauge, the shell coordinate time $t$ is identified with the proper time  $\tau$ of the  bulk observer. Without pointing out explicitly, we will always insist this gauge in the following.

For the wormhole spacetime to be considered in this work, the general form of the metric can be written as
\bea{StaMetAnsa}
ds^2=g_{MN}dx^Mdx^N=-A(r)dT^2 +B(r)dr^2 +R^2(r)d \Omega^2_2
\label{bulkMetric}
\eea
Since the thin shell is located at $r=r(\tau)$, the above metric is valid on $r(\tau)<r$ and its $Z_2$ symmetric region.  In this bulk space-time, an evolving thin shell will get a time dependent induced metric
\bea{}
&&\hspace{-5mm}ds^2=h_{\mu \nu} dx^\mu dx^\nu=-d\tau^2 +R^2[r(\tau)] d\Omega^2_2
\label{FLRWbrane}
\\
&&\hspace{-5mm}
h_{\mu\nu}=e^A _\mu e^B _\mu g_{AB}
\label{induceMetric}
\eea
From the bulk space observer's perspective, the induced metric on $\partial\Sigma$ could be given by the tangential components of the bulk metric \eqref{bulkMetric} with the projection tensor $h_{MN}=g_{MN}-n_M n_N$, where $n_M$ is the unit normal into the removed region $\Sigma_\pm$, i.e. to the $r<r(\tau)$ region. Orthogonalities and the normalization $u^M n_M =0, n^M n_M=1$ implies $n_{M}=\{(AB)^\frac{1}{2}\dot{r},-[B(1+B\dot{r}^{2})]^\frac{1}{2},\vec{0}\}$. Denoting the evolving speed of the thin shell in the bulk spacetime as $u^M=\{\dot{T}(\tau),\dot{r}(\tau),\vec{0}\}$, then the normalizing condition $u^M u_M=-1$ will tells us $\dot{T}(\tau)=[(1+B\dot{r}^{2})/A]^\frac{1}{2}$. 

Dynamics of the bulk-shell system is determined by the action principle
\bea{}
\label{Taction}
&&S=S_{bulk}+S_{shell}+S_{surf},~n=3+1
\\
&&S_{bulk}=\frac{1}{2\kappa ^2 _n} \int _M d^n x \sqrt{-g} [\mathcal{R}+\mathcal{L}_{M}]
\label{Bulkaction}
\\
&&S_{surf}=-\int_{\partial\Sigma_\pm}d^{n-1}x\sqrt{-h}[\frac{1}{\kappa_{n}^{2}}\mathcal{K}^{\pm}]
\label{GHterm}
\\
\label{Braneaction}
&&S_{shell}=\int_{\partial\Sigma_\pm}d^{n-1}x\sqrt{-h}\mathcal{L}_{shell}
\eea
where $S_{surf}$ is a Gibbons-Hawking term on $\partial\Sigma _\pm$ \cite{Gibbons:1976ue}, it makes the derivative of the metric across $\partial\Sigma$ continuous. The extrinsic curvature of $\partial\Sigma _\pm$ is defined as 
\bea{}
&&\hspace{-5mm}\mathcal{K}^\pm _{MN}=\frac{1}{2}h^P _M h^Q _N (\nabla_P n_Q+\nabla_Q n_P) \vert _\pm
\\
&&\hspace{-5mm}\mathcal{K}^\pm =h^{MN} K^\pm _{MN}
\eea
Varying $\eqref{Taction}$ respect to the metric, besides a bulk term which yields the standard Einstein equation, we obtain also a junction term on the boundary $\partial \Sigma_{\pm}$,
\bea{varyonboun}
\int_{\Sigma_\pm} \sqrt{-h} d^{n-1}x  ~[-D^{N}(h_{N}^{P}n^{Q}\delta g_{PQ})-\mathcal{K}_{MN}\delta g^{MN}+h_{MN}\mathcal{K}\delta g^{MN} -\kappa ^2 _n \tau _{MN} \delta g^{MN}]
\eea
with $D_M$ is the covariant derivative associating with the induced metric on $\Sigma _\pm$. By tangent projection methods $D_M=h_{M}^{K}\nabla_{K}$. For this reason, the first term in this variation is a total derivative and could be integrated away. As results, the junction condition on $\partial \Sigma$ can be written as,
\bea{JCbyBulk}
\{\mathcal{K}_{MN}-\mathcal{K}h_{MN}\}\vert_{\Sigma_+}-\{\mathcal{K}_{MN}-\mathcal{K}h_{MN}\}\vert_{\Sigma_-}=-\kappa ^2 _n \tau _{MN}
\eea
with $\tau _{MN}=e^\mu _M e^\nu _N \tau_{\mu \nu}$ and $\tau _{\mu \nu}=-2\delta \mathcal{L} _{shell} / \delta h^{\mu\nu}+h_{\mu\nu} \mathcal{L}_{shell}$ being the $2+1$-dimensional effective energy-momentum tensor contributed by shell matters. The motion trajectory $r(\tau)$ of the shell is determined by this junction condition. Considering the $Z_2$ symmetry, $\{\mathcal{K}_{MN}\}\vert_{\Sigma_-}=-\{\mathcal{K}_{MN}\}\vert_{\Sigma_+}$, this condition simplifies to
\bea{JCbyBulk}
\{\mathcal{K}_{MN}-\mathcal{K}h_{MN}\}\vert_{\Sigma_+}=-\frac{\kappa ^2 _n}{2} \tau _{MN}
\eea
For expression clearness, we will through away the subscript $+$ of $\Sigma_+$ in the following. By the bulk-shell viewpoint translation tool $e^\mu _A$ and its inverse $e^{A}_\mu$, the junction condition $\eqref{JCbyBulk}$ can also be written through internal coordinates of the shell,
\bea{JCbyBrane}
K_{\mu \nu}-K h_{\mu \nu}=-\frac{\kappa ^2 _n}{2} \tau _{\mu \nu}
\eea 
with $K_{\mu \nu}=e^M _\mu e^N _\nu \mathcal{K}_{MN}$. Alternatively, $\eqref{JCbyBrane}$ can also be written as
\bea{easiJCbyBrane}
K_{\mu \nu}=-\frac{\kappa^2 _n}{2}(\tau _{\mu \nu} - \frac{1}{n-2}\tau h_{\mu \nu}),~n=4
\eea
For our wormhole metric ansatz $\eqref{StaMetAnsa}$, the components of $K_{\mu \nu}$ are given by
\bea{}
\label{Kij}
K_{ij}=-\frac{\sqrt{1+B\dot{r}^{2}}}{\sqrt{B}} RR' \delta_{ij}\\
\label{Ktautau}
K_{\tau \tau}=\frac{1}{\sqrt{AB}}\frac{d}{dr}(\sqrt{A+AB\dot{r}^{2}}) 
\eea
 
\subsection{The building-block spacetime for wormholes}

In this paper, three types of 4D spherically symmetric static blackhole will be considered as the building-block spacetimes for constructing TSW. They are the central-singular AdS-, dS- Schwarzschild blackholes and central-regular Hayward one. All these three has simple line element of the form
\bea{}
\label{GeMeAnsatz}
&&ds^{2}=-f(r)dt^{2}+{f(r)}^{-1}dr^{2}+r^{2}d\Omega^{2}
\eea
with $d\Omega^{2}=d\theta^{2}+\sin^{2}\theta d\varphi^{2}$. We will set the gravitational coupling constant $G$ to be 1. The first two are characterized by  horizon functions
\bea{}
f(r)=1-\frac{2m}{r}\pm\frac{r^2}{\ell^2}
\label{dSAdSMetric}
\eea
The AdS-Schwarzschild space-time has single horizon determined by the mass parameter $m$. While the dS-Schwarzschild space-time has two horizons, with the inner one being the usual blackhole horizon while the outer one being a cosmological horizon. Both this two building-block spacetime has essential singularities at the $r=0$ point.
\begin{figure}[ht!]
\begin{center}
\includegraphics[scale=0.38]{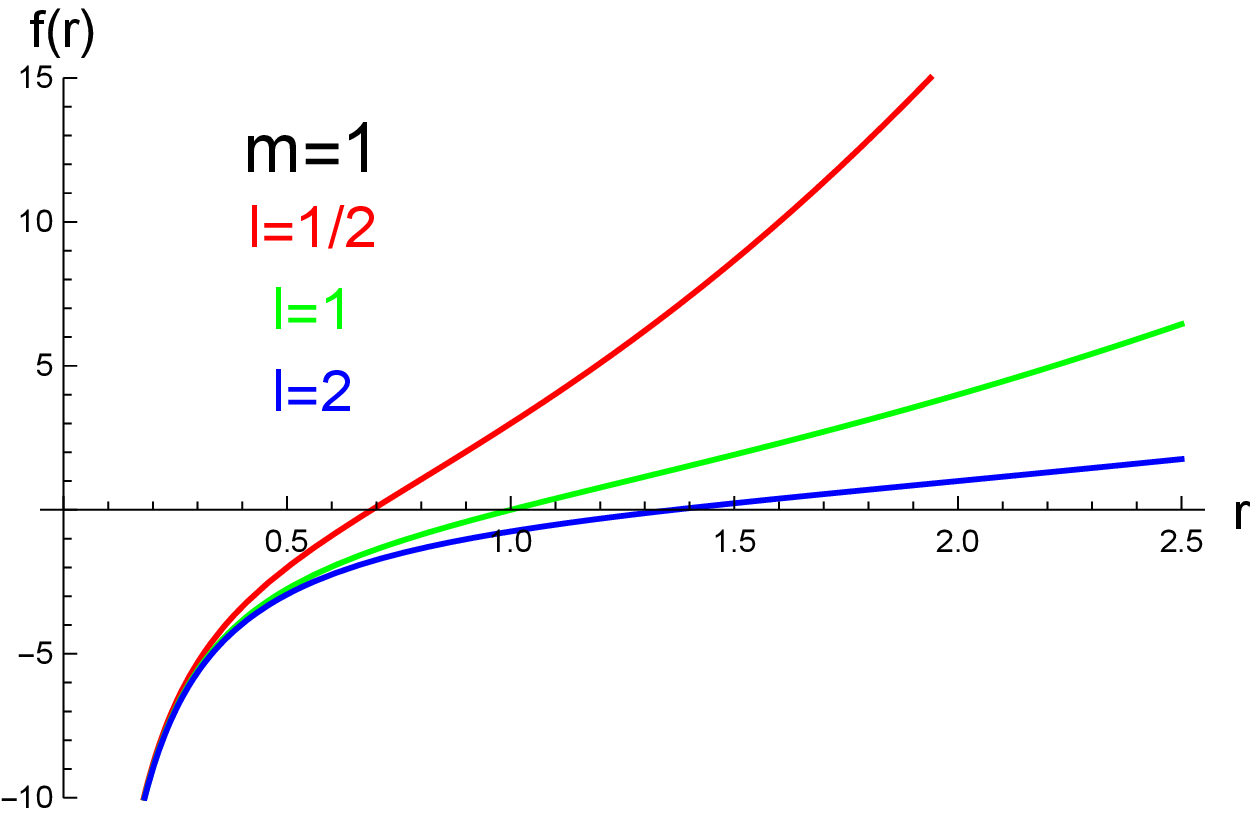}
\includegraphics[scale=0.38]{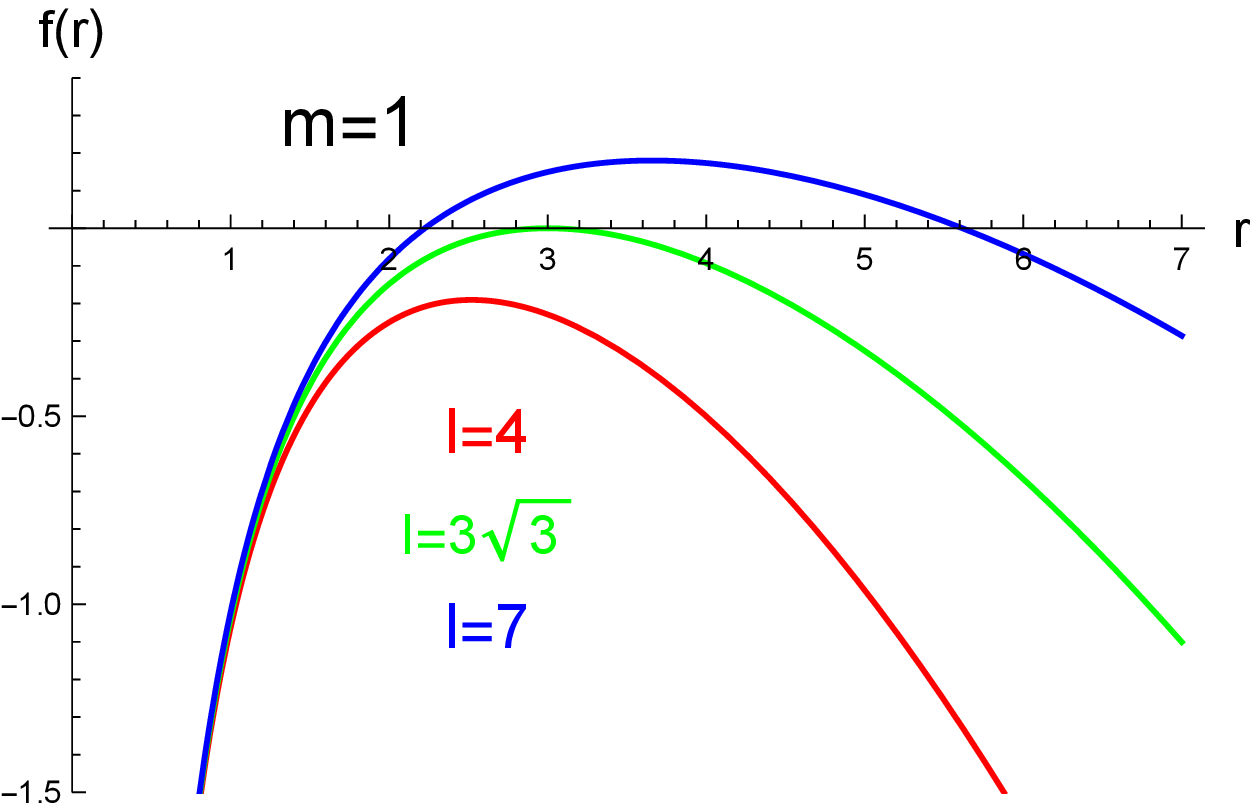}
\includegraphics[scale=0.38]{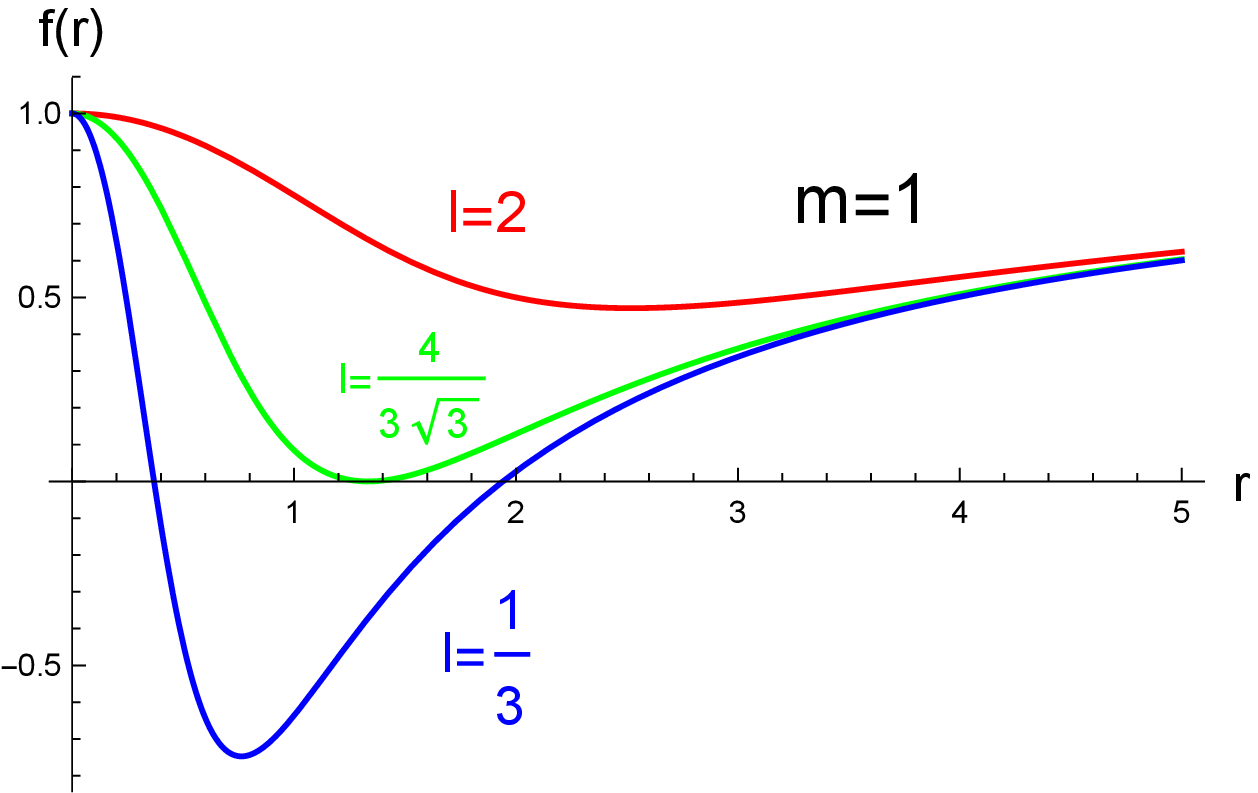}
\caption{(color online). The horizon function $f(r)$ of the AdS-, dS-Schwarzschild and Hayward blackhole space-time, each with three typical values of $\lambda\equiv\frac{\ell}{m}$.}
\label{figHorizon}
\end{center}
\end{figure}

The third building-block spacetime is the asymptotically flat but central regular Hayward blackhole characterized by the horizon function \cite{Frolov:2017dwy, Frolov:2017rjz}
\bea{}
\label{HaMetric}
f(r)=1-\frac{2mr^{2}}{r^{3}+2ml^{2}}
\eea 
whose asymptotics read
\bea{}
f(r)=\left\{
\begin{array}{l}
 1-\frac{2m}{r}+O(\frac{1}{r^4}),~r\to \infty
 \\
 1-\frac{r^2}{l^2}+O(r^5),~~r\to 0
\end{array}
\right.
\eea
From this asymptotic behavior, we see that the Hayward blackhole also has two horizons, with the outer a Schwarzschild-like one while the inner a de-Sitter like one. The Hayward metric $\eqref{HaMetric}$ follows from a nonlinear electrodynamics coupled with Einstein gravitation
\bea{EM}
\label{TotalEM}
&&S=\frac{1}{2}\int_{M}d^{4}x\sqrt{-g}[\mathcal{R}-\mathcal{L} (F)]\\
\label{NonLinMg}
&&\mathcal{L} (F)=-\frac{6}{l^{2}[1+(\frac{\beta}{F})^{3/4}]^{2}}
\eea
with the corresponding electromagnetic field strength given by
\bea{monopole}
F_{\theta \varphi}=P \sin \theta,~
\mathrm{with}~\beta = \frac{2P^2}{(2ml^2)^{4/3}}
\eea

Fig \ref{figHorizon} displays the three types of $f$ explicitly each with three typical values of $\lambda\equiv\frac{\ell}{m}$.

\subsection{Evolution equations of the TSW }
\label{DerEqMo}

Now, by the building-block spacetimes above, and using methods introduced in the preface of section \ref{MEofTSW}, we can easily construct three types of thin shell wormholes, with whose joining shell $\partial \Sigma$'s evolution controlled by $\eqref{easiJCbyBrane}$. Setting the energy-momentum tensor in the $\partial \Sigma$ as simple ideal fluid types,
\bea{Perfluid}
\tau ^\mu _\nu =Diag \{-\rho,P,P \}
\eea
the two independent components of the equation of motion will become
\bea{} 
&&\frac{d}{dr}(\sqrt{f+\dot{r}^{2}})=\frac{1}{4}\rho+\frac{1}{2}P 
\label{Ktautaueq}
\\
&&-\frac{1}{r}\sqrt{f+\dot{r}^{2}}=\frac{1}{4}\rho
\label{Kijeq}
\eea
Substituting $\eqref{Kijeq}$ into $\eqref{Ktautaueq}$, we get the more directive continuity equation 
\bea{}
\label{ConEq}
r\frac{d\rho}{d\tau}+2\rho \dot{r}+2\dot{r}P=0
\eea
This equation can also be derived out from the energy-momentum conservation law $D_\mu \tau ^{\mu \nu}=0$, with $D_\mu$ being the covariant derivative associating with $h_{\mu \nu}$ on $\partial \Sigma$.

Obviously, evolution processes of the wormhole throat are described by $r(\tau)$. Given equation of states of the thin shell filling matters
\bea{stateEQ}
P=\phi (\rho)
\eea
we can solve out $\rho(r)$ from eq.$\eqref{ConEq}$ to obtain 
\bea{solrhor}
\frac{dr}{r}=-\frac{1}{2}\frac{d\rho}{\rho+\phi(\rho)}
\eea
As long as $\rho(r)$ is known, we can substitute it into \eqref{Kijeq} and write the wormhole throat's equation of motion into a Friedmann like equation
\bea{}
\dot{r}^{2}=H(r),~
H(r)=\frac{\rho^{2}(r) r^2}{16}-f(r)
\label{GeMoPoForTSW}
\eea
Meaningful solutions to this equation exists only when $0\leqslant H(r)$. While previous works on TSW's construction and stability analysis exclusively focus on parameter choices so that $H(r)<0$. So their TSW are born-static which allows no true dynamic evolutions at all.

\subsubsection{Shells inhabiting phantom gas}

\begin{figure}[!ht]
\begin{center}
\includegraphics[scale=0.45]{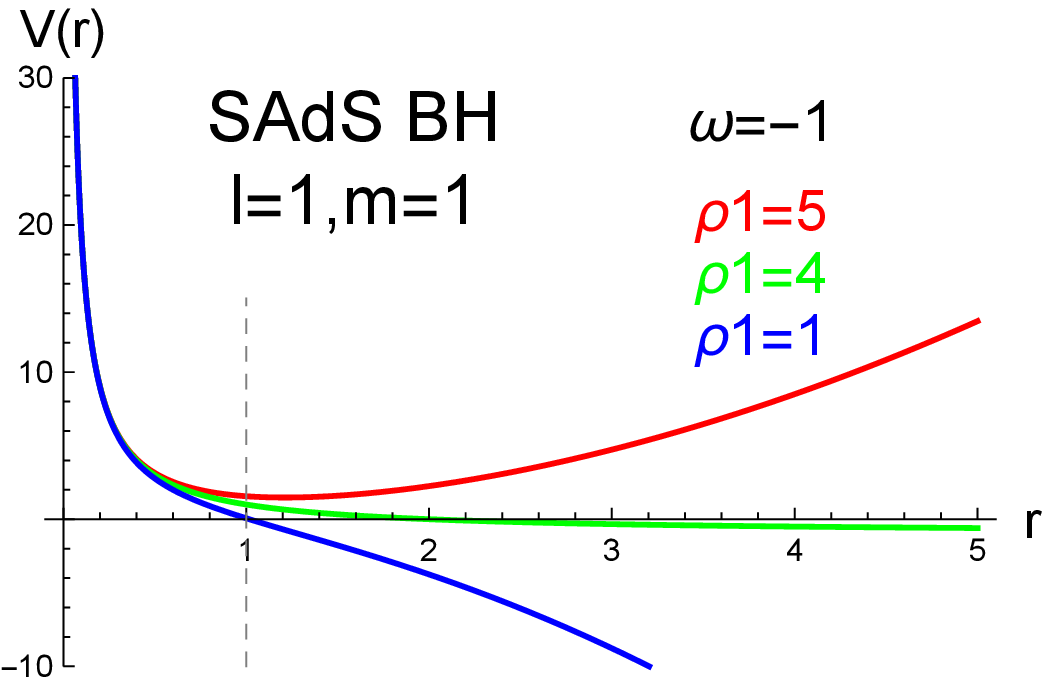}
\includegraphics[scale=0.45]{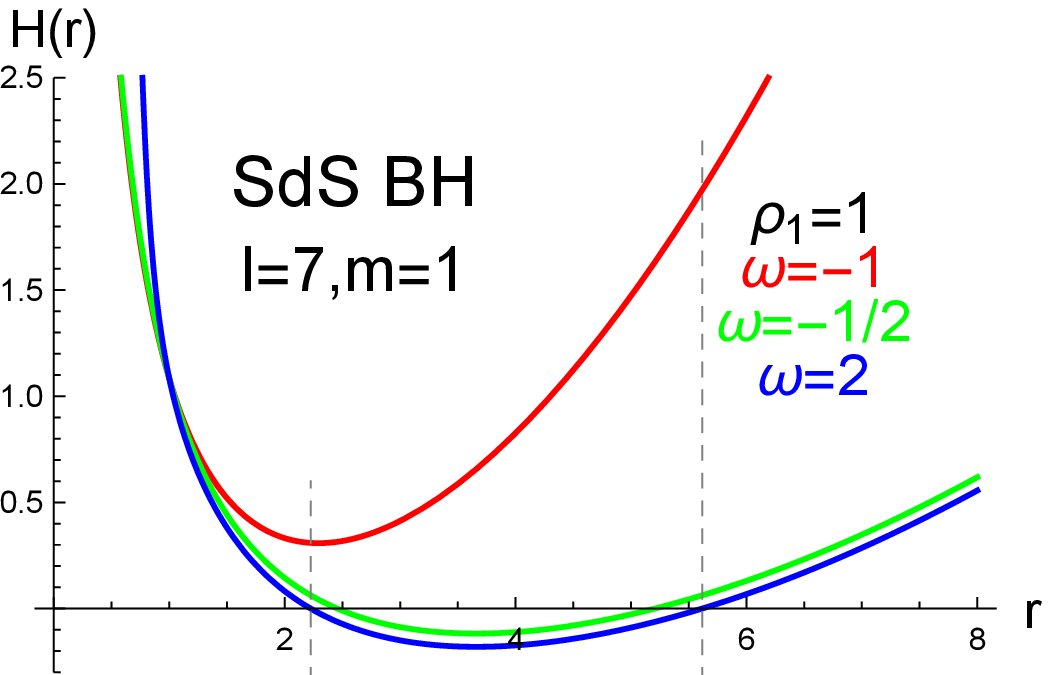}
\\
\includegraphics[scale=0.45]{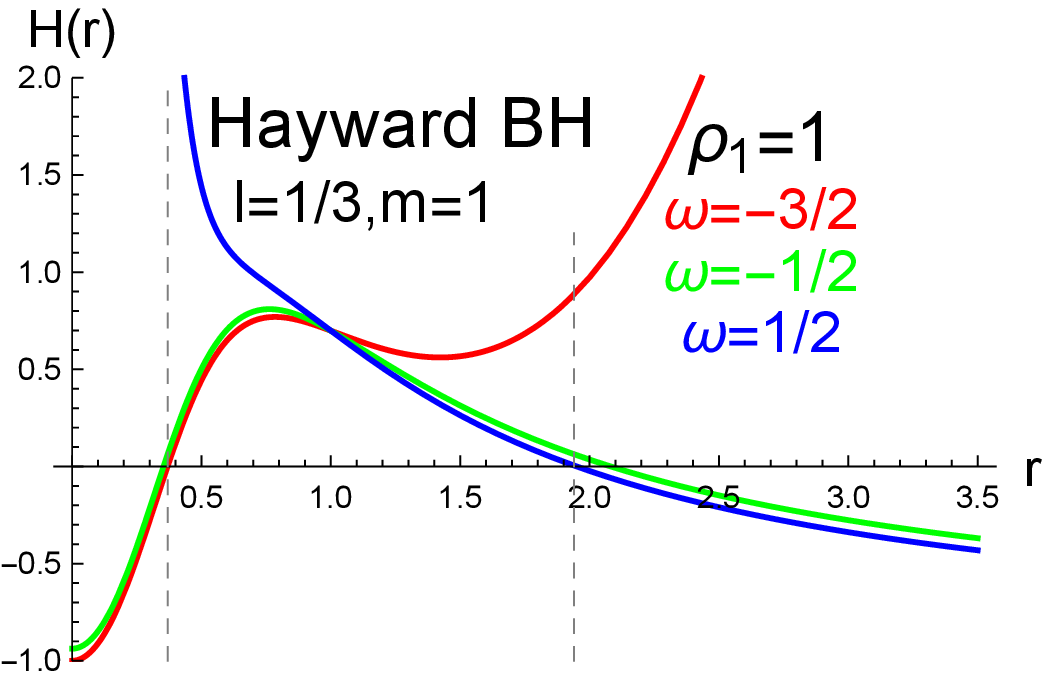}
\includegraphics[scale=0.45]{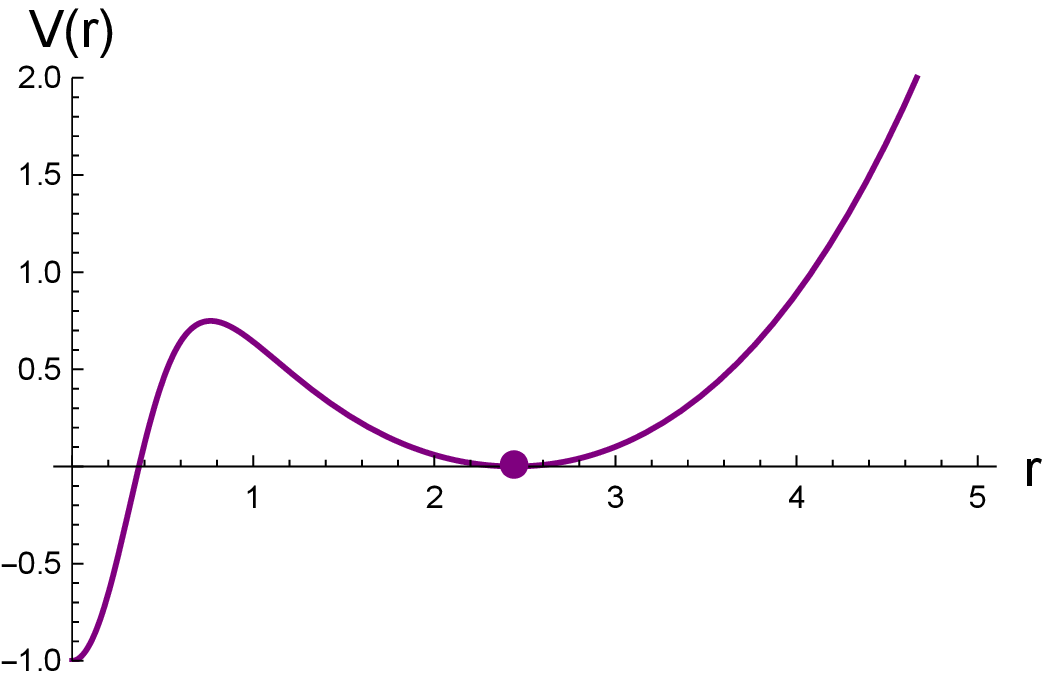}
\caption{(color online) The throat evolving function $\dot{r}^2=H(r)$ of the wormholes constructed from three blackhole space-times with LGs living on the joining shell. The bottom-right panel is a representative $H(r)$ line along which an equilibrium point exists, $H=H'=0$ at this point.}
\label{figTSWtraLGB}
\end{center}
\end{figure}

We will consider two types of matters filling on the thin shell. The first is the phantom gas (PG). While the second is the Chaplygin gas (CG). For the former, $P=\omega \rho$. We are mainly interesting in $\omega<-1$ coefficient. But in some cases, we will consider $-1<\omega$ for special purposed variation trends illustration. Obviously, for PG eq.$\eqref{solrhor}$ can be integrated analytically, with results
\bea{}
\rho=\rho_{1}r^{-2(1+\omega)}
\label{RhoVrLGB}
\eea
Substituting this relation into $\eqref{GeMoPoForTSW}$ and considering differences between the building-block spacetime metric $\eqref{dSAdSMetric}-\eqref{HaMetric}$, we have
\bea{MoEqPoOfthoLBG}
H(r)=\left\{
\begin{array}{l}
\frac{\rho_{1}^{2}}{16}~r^{2-4(1+\omega)}-1+\frac{2m}{r}+\frac{r^2}{l^2}~,~~(SdS)
\\
\frac{\rho_{1}^{2}}{16}~r^{2-4(1+\omega)}-1+\frac{2m}{r}-\frac{r^2}{l^2}~,~~(SAdS)
\\
\frac{\rho_{1}^{2}}{16}~r^{2-4(1+\omega)}-1+\frac{2mr^{2}}{r^{3}+2ml^{2}}~,~~(Hayward)
\end{array}
\right.
\label{VrLBG}
\eea
Figure.\ref{figTSWtraLGB} displays this function $H(r)$ with some typical $\omega$ and $\rho_1$ explicitly. From the figure and eq.\eqref{GeMoPoForTSW}, we know that in the Hayward wormhole case, three kinds of motion modes are possible. Following ref.\cite{Chamblin:1999ya}, we will call them recollapsing phase (blue curve), bouncing phase (green curve) and expanding phase (red curve). Obviously, in the AdS- and dS-Schwarzschild wormholes, no bouncing phase exists.
According to the cut-and-paste method, only $r_h<r(\tau)$ region belongs to the wormhole space-time. This means that what really interests us is the expanding phase with regions outside the building-blocks' horizon. From eq.\eqref{VrLBG}, we can show that to obtain an expanding phase, the parameters $\omega, \rho_1$ are restricted by: (i) $\omega=-1 \& \rho_1 > \frac{4}{l}$ or $\omega<-1$ in the SAdS case, (ii) $\omega, \rho_1$ are arbitrary for SdS case and (iii) $\omega=-\frac{1}{2} \& \rho_1 \ge 4$ or $\omega<-\frac{1}{2}$. 
\begin{figure}[!ht]
\begin{center}
\includegraphics[scale=0.45]{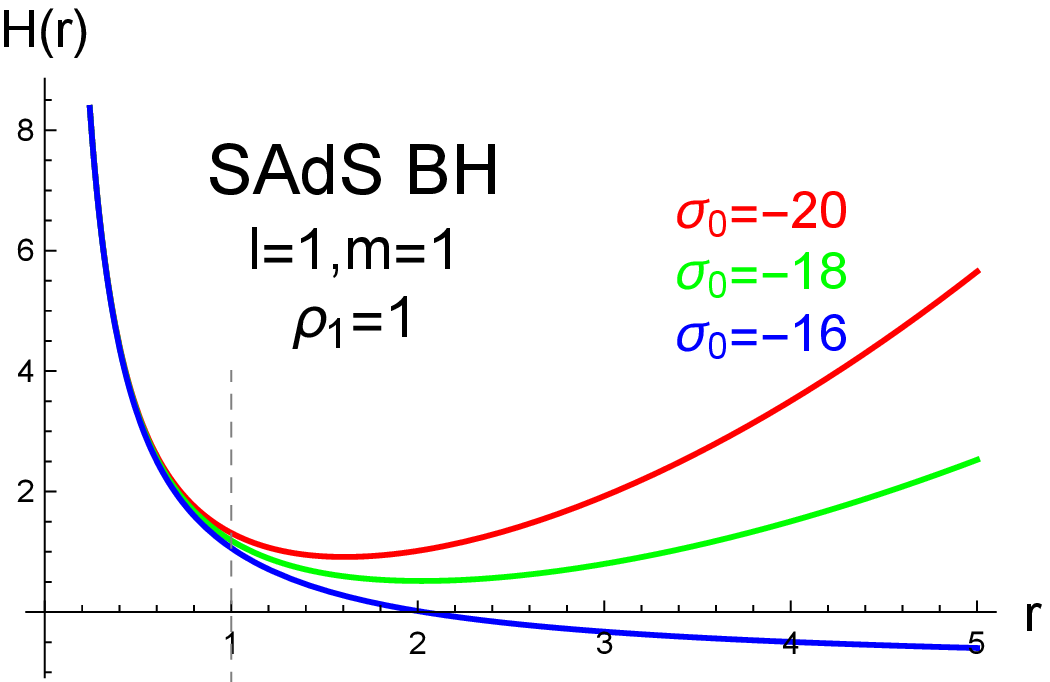}
\includegraphics[scale=0.45]{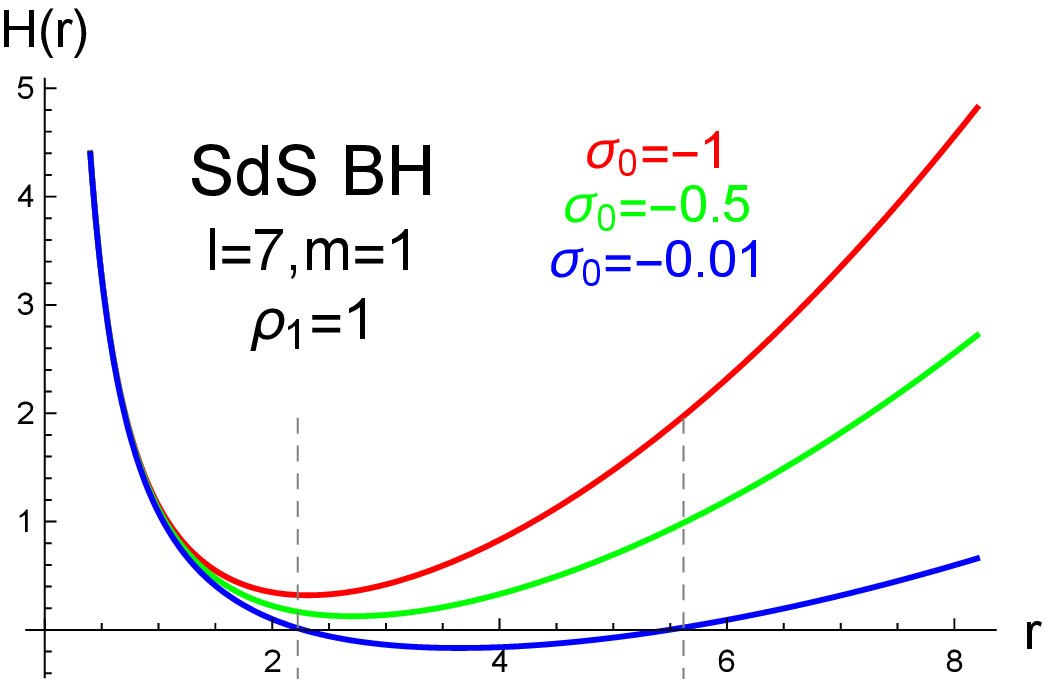}
\\
\includegraphics[scale=0.45]{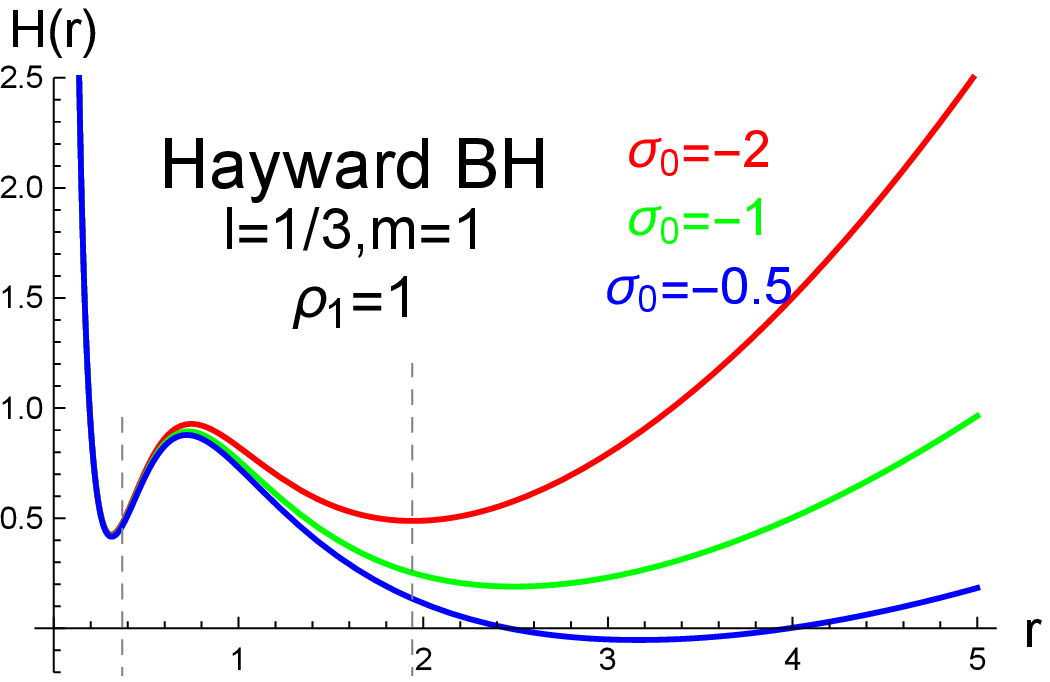}
\includegraphics[scale=0.45]{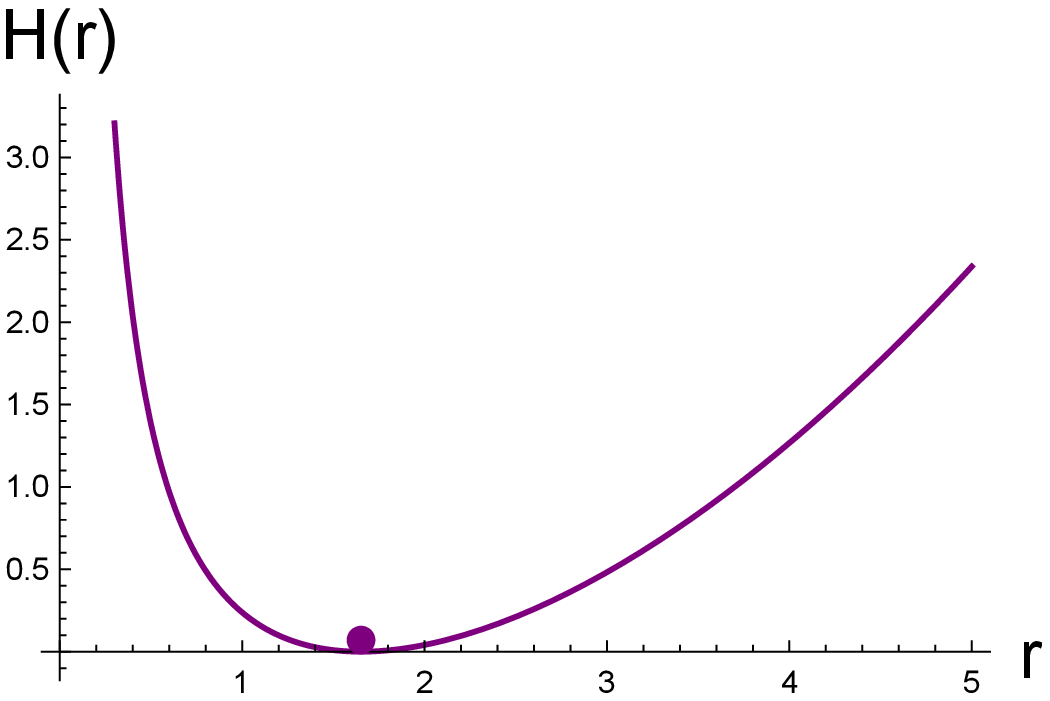}
\caption{(color online) Similar as figure \ref{figTSWtraLGB} but with CGs living on the joining shell.}
\label{figTSWtraCG}
\end{center}
\end{figure}

\subsubsection{shells inhabiting Chaplygin gas}

The second matter supporting the wormhole throat we will consider is the Chaplygin gas (CG). This is an exotic matter considered widely in cosmologies, due to its potential to give the accelerating expansion of the late universe phenomenological explanations. Traversable wormholes supporting by such gases are also considered in Ref.\cite{Lobo:2005vc,Eiroa:2007qz}. So, it is very natural to see what kind of throat evolutions $r(\tau)$ will be induced in the wormholes following from the cut-and-paste procedure. The equation of state representing the Chaplygin gas has the form
\bea{CGstate}
P\rho =\sigma _0~,~\sigma_0 <0
\eea
Substituting into eq.$\eqref{solrhor}$, the resulting energy density $\rho(r)$ can also be solved out analytically
\bea{}
\rho (r)=\sqrt{\frac{\rho_1}{r^4}-\sigma_0}
\eea
with $\rho_1$ another integration constant similar to that in eq.\eqref{RhoVrLGB}. We will assume $0<\rho_1$ to ensure $\rho(r)$'s well definition everywhere. For this $\rho(r)$, the potential function controlling the wormhole throats' evolution has the form
\bea{MoEqPoOfthoLBG}
H(r)=\left\{
\begin{array}{l}
\frac{\rho_1}{16 r^2}-\frac{\sigma _0 r^2}{16}-1+\frac{2m}{r}+\frac{r^2}{l^2}	~~,~(SdS)
\\
\frac{\rho_1}{16 r^2}-\frac{\sigma _0 r^2}{16}-1+\frac{2m}{r}-\frac{r^2}{l^2}~~,~(SAdS)
\\
\frac{\rho_1}{16 r^2}-\frac{\sigma _0 r^2}{16}-1+\frac{2mr^{2}}{r^{3}+2ml^{2}}~~,~(Hayward)
\end{array}
\right.
\eea
Figure.\ref{figTSWtraCG} displays this $H(r)$ for some typical $\sigma_0$ or $\rho_1$s intuitively. From the figure and eq.\eqref{GeMoPoForTSW}, we easily see that to obtain an expanding wormhole throat, in the large $r$ region, $0<H(r)$ must be fulfilled. However, in all three building-block spacetimes, parameter choices are always possible to fullfil this condition.

\subsection{Stability analysis around the equilibrium point}

The perturbation method for stability analysis of spherically symmetric thin-shell wormholes is introduced by \cite{Poisson:1995sv}. In this method, we first find a static wormhole structure. This means that the function $H(r)$ introduced in the previous section has an equilibrium point $r_0$ on which $H(r_0)=H'(r_0)=0$, correspondly $\dot{r}_0=\ddot{r}_0=0$. Just as we plotted in the bottom-right panel of figure.\ref{figTSWtraLGB} and \ref{figTSWtraCG}, by adjusting the equation of state and the integration constant featuring matters living on the joining shell, it  is always possible to find such $H$-$r$ curves with equilibrium point $r_0$ in all three building-block spacetimes. Starting from eq.\eqref{GeMoPoForTSW}, ref.\cite{Poisson:1995sv} considers a linearized radial perturbation of the wormhole throat
\bea{}
[\delta \dot{r}(\tau)]^2=H[r_0+\delta r (\tau)]-H[r_0]
=\frac{1}{2}H''[r_0][\delta r(\tau)]^2+O[\delta r(\tau)^3]
\label{clStabAnalysis}
\eea
and get conclusions that, the wormhole is stable as long as the corresponding $H''(r_0)<0$, since the perturbation quantity $\delta{r}(\tau)$ in this case has oscillation solution. Otherwise, the perturbation will grow with time, the corresponding wormhole is unstable. By this idea, ref.\cite{Poisson:1995sv} makes general stability analysis without particular equations of state specified on the wormhole throat, they uncover regions of stability in the $(\beta^2 _0)-(r_0/M)$ plane, with $\beta_0=\frac{\partial P/\partial r}{\partial \rho/\partial r}\vert_{r_0}$ defined as the sound speed of matters filling in the joining shell. As complements, ref.\cite{Lobo:2003xd} considers the stability question of TSWs constructed on building-block spacetimes with nonzero cosmological constants $\Lambda$. It is found that the region of stability is significantly extended in the positive $\Lambda$ case. Ref.\cite{Lobo:2005yv} makes analysis for wormholes with the joining shell filled with phantom-like matters defined by $P=\omega \rho~,~\omega<-1$. Ref.\cite{Eiroa:2007qz} figures out the region of stability in $(\tilde{A}M^2)-(r_0/M)$ plane for the chaplygin gas filled wormholes, where $\tilde{A}$ is positively related with the equations of state $P\rho=-A$, and the building-block spacetime covers Schwarzschild, Schwarzschild-de Sitter, Schwarzschild-anti-de Sitter Reissner-Nordstrom et al. Stability analysis are also made for wormholes building from more interesting building-bolck spacetimes, such as those featured by non-spherical topology \cite{Lemos:2008aj, Sharif:2013efa}, noncommutative geometry \cite{Garattini:2008xz}, $f(R)$ modified gravity \cite{Lobo:2009ip}, nonminimal curvature-matter coupling \cite{Garcia:2010xb}, and equation of state supporting the joining shell such as quark matter at ultra-high densities \cite{Harko:2014oua}, generalized Chaplin gas \cite{Gorini:2008zj}, $log$-form equation of state gas \cite{Halilsoy:2013iza}.

However, just as we pointed out in the preface of this section, all these works judge stabilities by criteria $H_{r_0}=H'_{r_0}=0$ and $H''_{r_0}<0$. Around such point, $H\leqslant0$,  so the joining shell are born-static. They cannot experience any real evolution since $\dot{r}^2=H<0$. This is unnatural and further considerations are necessary. Simple numeric works indicate that, there exist two different kinds of equilibrium points in the $H-r$ plane, just as we marked by $r_{0,\mathrm{ref}}$ and $r_{0,\mathrm{our}}$ in figure \ref{figEPComRef}. Around $r_{0,\mathrm{our}}$, $H=H'=0$, $0<H''$, the joining shell can experience real evolutions and reach point $r_0$. As long as perturbations to the system are stable, we expect that such configurations should also lead to stable wormholes. Of course, in this case, we need to make stability analysis on an evolving thin shell instead a born-static point.

\section{Evolving TSWs' stability to perturbations}
\label{PerofETSW}

\begin{figure}[!ht]
\begin{center}
\includegraphics[scale=1]{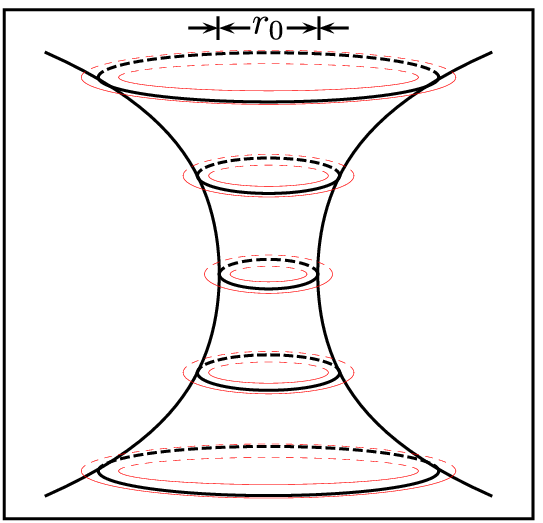}
\includegraphics[scale=1]{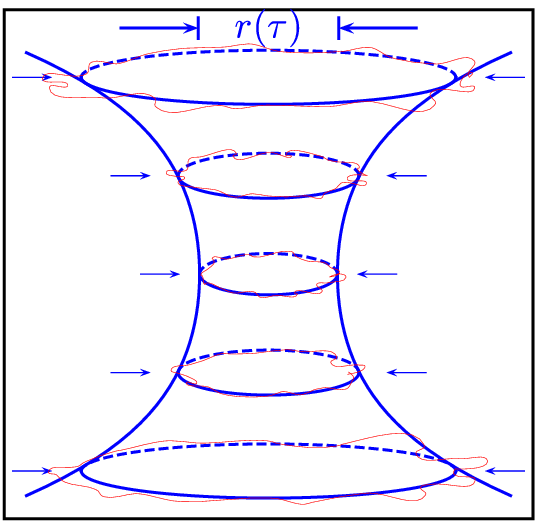}
\caption{(color online)  The left panel is radial perturbations to a born-static wormhole throat geometry. While the right panel is general perturbations to a wormhole throat in dynamical evolutions.}
\label{figpert}
\end{center}
\end{figure}

We easily see that perturbations in eq.\eqref{clStabAnalysis} are spheric symmetric. In practices, matter fluctuations on the joining shell $\partial\Sigma$ cannot be so symmetric. Quantum mechanically, the inhomogeneous and anisotropy fluctuations are unavoidable. If these fluctuations grows drastically as time passes by, the wormhole throat will be teared up. In such cases, the wormhole is unstable to perturbations, otherwise, the wormhole is stable. Figure \ref{figpert} compares the spherical symmetric perturbation to a static wormhole throat and general perturbations to an evolving wormhole throat schematically.

\subsection{Fluctuations of the wormhole throat with unperturbed bulk}

The purpose of this section is to build the theory of perturbations' evolution on the wormhole throat. The main idea and formula are borrowed from cosmological perturbation theories, especially the brane world cosmology \cite{Garriga:1991ts,Guven:1993ew,Boehm:2002kf,Brax:2002nx}. The final result is a differential equation controlling the time evolution of matter fluctuations on the wormhole throat $\partial \Sigma$. 

We will use the barred symbol to denote the unperturbed classical quantities introduced in previous sections, while those without bars denoting the perturbed ones. In terms of embedding coordinates in the bulk spacetime, fluctuations of the wormhole throat can be written as 
\bea{FluofCoorPosi}
X^{A}(x^{\mu})=\bar{X}^{A}(x^{\mu})+\zeta(x)\bar{n}^{A}
\eea
From eq.\eqref{induceMetric}, we can get the induced metric on the wormhole throat through $h_{\mu\nu}=e^M _\mu e^N _\nu g_{MN}$, which can be written as
\bea{}
h_{\mu\nu}=\bar{h}_{\mu\nu}+2\zeta\bar{\mathcal{K}}_{\mu\nu}
,~h^{\mu\nu}=\bar{h}{}^{\mu\nu}-2\zeta\bar{\mathcal{K}}{}^{\mu\nu}
\eea 
A projection trick $\partial_{\nu}\bar{n}^{N}=\frac{\partial\bar{X}^{A}}{\partial x^{\nu}}\nabla_{A}\bar{n}^{N}$ has been used here and the definition of $\mathcal{K}_{\mu\nu}$ is as follows
\beq{}
\mathcal{K}_{\mu \nu}=\frac{1}{2}e^M _\mu e^N_\nu (\nabla _M n_N+\nabla _N n_M)
\label{calKdefinition}
\eeq

Considering fluctuations $\eqref{FluofCoorPosi}$, the normal vector of the wormhole throat can be written as
\bea{deltan}
n^{A}=\bar{n}^{A}+\alpha\bar{n}^{A}+\beta^{\mu}\frac{\partial\bar{X}^{A}}{\partial x^{\mu}}
\eea 
$\alpha$ and $\beta^{\mu}$ here can be determined through the leading order definition/normalization equations $n^Ae_A=0$, $n^An^Bg_{AB}=1$
\bea{}
&& \alpha=-\frac{1}{2}\zeta\bar{n}^{M}\bar{n}^{N}\bar{n}^{C}\partial_{C}\bar{g}_{MN}\\
&& \beta^\mu=-\bar{h}^{\mu \nu}(\partial_{\nu}\zeta+\zeta\bar{e}_{\nu}^{B}\bar{n}^{C}\bar{n}^{A}\partial_{C}\bar{g}_{AB})
\eea
As results ($\delta n_b\equiv n-\bar{n}$)
\bea{}
\delta n_{B}=-\frac{1}{2}\zeta\bar{n}_{B}\bar{n}^{C}\bar{n}^{M}\bar{n}^{N}\partial_{C}\bar{g}_{MN}-\bar{g}_{BA}\bar{h}{}^{\mu\nu}\bar{e}_{\mu}^{A}\partial_{\nu}\zeta
\eea
Further more, 
\bea{}
&&\nabla_{A}n_{B}=\nabla_{\bar{A}}\bar{n}_{B}+\partial_{A}\delta n_{B}-\bar{\Gamma}{}_{AB}^{M}\delta n_{M}-\frac{1}{2}(\bar{g}^{MN}\bar{n}_{M}\delta\Gamma_{NAB}+\bar{n}_{M}\delta g^{MN}\bar{\Gamma}_{NAB})\\
\nonumber
&&\delta\Gamma_{NAB}=\partial_{A}(\zeta\bar{n}^{C}\partial_{C}\bar{g}_{BN})+\partial_{B}(\zeta\bar{n}^{C}\partial_{C}\bar{g}_{NA})-\partial_{N}(\zeta\bar{n}^{C}\partial_{C}\bar{g}_{AB})\\
\nonumber
&&\delta g^{MN}=-\zeta\bar{g}^{MP}\bar{g}^{NQ}\bar{n}^{C}\partial_{C}\bar{g}{}_{PQ}
\eea
With these preparation, the perturbed extrinsic curvature \eqref{calKdefinition} can be calculated routinely. To leading order in $\zeta$, the components are,
\bea{}
\nonumber
&&\hspace{-5mm}\delta\mathcal{K}_{\tau \tau}=-\ddot{\zeta}+\dot{\zeta}\bar{u}{}^{B}[\bar{n}{}^{A}\nabla_{\bar{A}}\bar{n}_{B}-\bar{n}^{C}\bar{n}^{N}\partial_{C}\bar{g}_{BN}]+\zeta\bar{g}^{AL}\bar{u}^{N}\bar{u}{}^{B}\nabla_{\bar{N}}\bar{n}_{L}(\nabla_{\bar{A}}\bar{n}_{B}+\nabla_{\bar{B}}\bar{n}_{A})\\
\nonumber
&&\hspace{-5mm}\quad~\quad \quad+\zeta\bar{u}{}^{A}\bar{u}{}^{B}[\bar{\Gamma}{}_{AB}^{M}\bar{n}_{M}\bar{n}^{Q}\bar{n}^{C}\partial_{C}\bar{n}_{Q}-\bar{n}^{C}\bar{n}^{M}\partial_{A}\bar{n}_{B}\partial_{C}\bar{n}_{M}-\frac{1}{2}\bar{n}^{N}\bar{n}^{C}\partial_{C}\bar{\Gamma}_{NAB}\\
\label{DelLowtata}
&&\hspace{-5mm}
\quad\quad\quad\quad\quad\quad~\quad -\frac{1}{2}\bar{n}_{Q}\bar{\Gamma}_{NAB}\bar{n}^{C}\partial_{C}\bar{g}^{NQ}-\bar{n}^{N}\partial_{C}\bar{g}_{BN}\partial_{A}\bar{n}^{C}+\frac{1}{2}\bar{n}^{N}\partial_{C}\bar{g}_{AB}\partial_{N}\bar{n}^{C}]\\
\label{DelLowtai}
&&\hspace{-5mm}\delta\mathcal{K}_{\tau i}=-\partial_{i}\dot{\zeta}+\partial_{i}\zeta[\frac{1}{2}\bar{n}^{A}\bar{u}^{B}\nabla_{\bar{A}}\bar{n}_{B}-\frac{1}{2}\bar{n}^{C}\bar{u}{}^{B}\bar{n}^{N}\partial_{C}\bar{g}_{BN}]+\frac{1}{2}\bar{u}{}^{B}\bar{h}^{kj}\partial_{B}\bar{g}_{ki}\partial_{j}\zeta\\
\nonumber
&&\hspace{-5mm}\delta\mathcal{K}_{ij}=-\partial_{i}\partial_{j}\zeta+\frac{1}{2}\dot{\zeta}\bar{u}^{N}\partial_{N}\bar{g}_{ij}+\frac{1}{2}\zeta[\bar{g}^{kl}\bar{n}^{Q}\bar{n}^{N}\partial_{Q}\bar{g}_{jl}\partial_{N}\bar{g}_{ki}-\bar{n}^{N}\bar{n}^{C}\bar{n}^{Q}\partial_{N}\bar{g}_{ij}\partial_{C}\bar{n}_{Q}\\
\label{DelLowij}
&&\hspace{-5mm}\quad\quad\quad\quad\quad\quad~ +\bar{n}^{N}\partial_{C}\bar{g}_{ij}\partial_{N}\bar{n}^{C}+\bar{n}^{C}\bar{n}^{N}\partial_{C}\partial_{N}\bar{g}_{ij} +\bar{n}^{C}\bar{n}_{Q}\partial_{N}\bar{g}_{ij}\partial_{C}\bar{g}^{NQ}]
\eea
For use in the following, the perturbed extrinsic tensor $\delta\mathcal{K}^\mu _\nu$ with mixed index should also be given here
\bea{}
\label{uptrastodown}
\delta\mathcal{K}{}_{\nu}^{\mu}=\bar{h}^{\mu\alpha}\delta\mathcal{K}_{\alpha\nu}-2\zeta\bar{\mathcal{K}}^{\mu\alpha}\bar{\mathcal{K}}_{\alpha\nu}
\eea
Up to now, to first order, all geometrical quantities describing fluctuations of the wormhole throat are obtained.

Now, let us consider the perturbed junction condition which relates matters and geometry between the two sides of the wormhole joining shell,
\bea{}
\label{perjunc}
\delta\mathcal{K}_{\nu}^{\mu}=-\frac{1}{2}(\delta\tau_{\nu}^{\mu}-\frac{1}{n-2}\delta\tau\delta_{\nu}^{\mu})~,~n=4
\eea
Imitating the standard 4D cosmological perturbation theory, we take the longitudinal gauge here and write components of the perturbed energy momentum tensor as follows
\bea{}
\label{peremtautau}
&& \delta\tau_{\tau}^{\tau}=-\delta\rho \\
\label{peremtaui}
&& \delta\tau_{i}^{\tau}=R(1+\omega)\rho\partial_{i}v \\
\label{peremij}
&& \delta\tau_{j}^{i}=\delta p\delta_{j}^{i}+\delta\pi_{j}^{i}
\eea
with the traceless anisotropic stress tensor $\delta\pi_{j}^{i}$ defined as
\bea{}
\delta\pi_{j}^{i}=\partial^{i}\partial_{j}\delta\pi-\frac{1}{n-2}\delta_{j}^{i}\partial^{k}\partial_{k}\delta\pi ~,~n=4
\eea
Substituting these matters' perturbation into eq.\eqref{perjunc}, we will obtain
\bea{}
\label{deltaMtt}
&& \delta\mathcal{K}_{\tau}^{\tau}=\frac{1}{2}(\frac{n-3}{n-2}\delta\rho+\delta P)~,~n=4\\
\label{deltaMti}
&&\delta\mathcal{K}_{i}^{\tau}=-\frac{1}{2}R(\rho+P)\partial_{i}v\\
\label{deltaMij}
&&\delta\mathcal{K}_{j}^{i}=\left\{
\begin{array}{ll}
-\frac{\kappa_{5}^{2}}{2}(\partial^{i}\partial_{j}\delta\pi) \quad (i\ne j)\\
-\frac{\kappa_{5}^{2}}{2}\delta\rho \quad (i=j~with~sum)
\end{array}
\right.
\eea

Equations \eqref{deltaMtt}-\eqref{deltaMij} and \eqref{DelLowtata}-\eqref{uptrastodown} forms the basic equation of motion controlling the evolution of $\zeta, \delta\rho, \delta P, v ,\delta\pi$ on the wormhole throat.
In our wormholes with (2+1)-dimensional thin joining-shell and asymptotics \eqref{GeMeAnsatz}, these equations become
\bea{}
&&\hspace{-5mm}\ddot{\zeta}-\frac{2f^{'}\dot{r}}{f^{2}}(f+\dot{r}^{2})\dot{\zeta}+\big[\frac{(f^{'})^{2}}{f}+\frac{5(f^{'})^{2}\dot{r}^{2}}{2f^{2}}+\frac{2(f^{'})^{2}\dot{r}^{4}}{f^{3}}
\\
&&\hspace{10mm}+\frac{1}{2}f^{''} -\frac{\dot{r}^{2}f^{''}}{2f}-\frac{\dot{r}^{4}f^{''}}{f^{2}}\big]\zeta=(\frac{1}{4}\delta\rho+\frac{1}{2}\delta P)
\nonumber
\eea
\beq{}
\partial_{i}\dot{\zeta}-\partial_{i}\zeta(\frac{\dot{r}}{r}+\frac{f^{'}\dot{r}}{f}+\frac{f^{'}\dot{r}^{3}}{f^{2}})=-\frac{1}{2}R(\rho+P)\partial_{i}v
\eeq
\beq{}
-\frac{1}{r^{2}}\partial^{i}\partial_{j}\zeta=-\frac{1}{2}\partial^{i}\partial_{j}\delta\pi\quad(i\ne j)
\eeq
\beq{}
-\frac{1}{r^{2}}\partial^{k}\partial_{k}\zeta+2\frac{\dot{r}}{r}\dot{\zeta}+2\zeta(\frac{f}{r^{2}}+\frac{2f^{'}}{r}+\frac{\dot{r}^{2}}{r^{2}}+\frac{5}{2}\frac{f^{'}}{f}\frac{\dot{r}^{2}}{r}+\frac{f^{'}}{f^{2}}\frac{\dot{r}^{4}}{r})=-\frac{1}{2}\delta\rho
\eeq
Focusing on scalar perturbation quantities $\zeta$, $\delta\rho$, $\delta P$ and write them into superpositions of all possible spherical harmonic modes
\bea{}
\zeta (\tau,\theta,\varphi)=\sum_{\ell m}\eta_{\ell m}(\tau)Y_{\ell m}(\theta,\varphi)
\eea
Their equations will become
\beq{}
\label{deltaDensity}
\Big\{-4\frac{\dot{r}}{r}\dot{\eta}-2\big[\frac{2f}{r^{2}}+\frac{4f^{'}}{r}+\frac{2\dot{r}^{2}}{r^{2}}+5\frac{f^{'}}{f}\frac{\dot{r}^{2}}{r}+\frac{2f^{'}}{f^{2}}\frac{\dot{r}^{4}}{r}+\frac{\ell(\ell+1)}{r^{2}}\big]\eta=\delta\rho
\Big\}_{\ell m}
\eeq
\bea{}
&&\hspace{-5mm}\Big\{2\ddot{\eta}+\big(2\frac{\dot{r}}{r}-\frac{4f^{'}\dot{r}}{f}-\frac{4f^{'}}{f^{2}}\dot{r}^{3})\dot{\eta}+[\frac{2(f^{'})^{2}}{f}+\frac{5(f^{'})^{2}\dot{r}^{2}}{f^{2}}+\frac{4(f^{'})^{2}\dot{r}^{4}}{f^{3}}+f^{''}-\frac{\dot{r}^{2}f^{''}}{f} 
\label{deltaPress}\\
&&\hspace{-5mm} -\frac{2\dot{r}^{4}f^{''}}{f^{2}}+\frac{2f}{r^{2}}+\frac{4f^{'}}{r}+\frac{2\dot{r}^{2}}{r^{2}}+5\frac{f^{'}}{f}\frac{\dot{r}^{2}}{r}+\frac{2f^{'}}{f^{2}}\frac{\dot{r}^{4}}{r}+\frac{\ell(\ell+1)}{r^{2}}\big]\eta
=\delta P\}_{\ell m}
\nonumber
\eea
Combining with \eqref{stateEQ} and $\delta P = \phi'(\rho) \delta \rho$, we obtain a final equation controlling the evolution of $\eta$ as follows
\beq{}
\ddot{\eta}_{\ell m}
+b\dot{\eta}_{\ell m}+c\eta_{\ell m}=0
\label{finalPerEQ}
\eeq
\beq{}
b=[2\phi'(\rho)+1]\frac{\dot{r}}{r}-\frac{2f^{'}\dot{r}}{f}-\frac{2f^{'}}{f^{2}}\dot{r}^{3}
\eeq
\bea{}
&&\hspace{-5mm}
c=(\frac{1}{f}+\frac{5\dot{r}^{2}}{2f^{2}}+\frac{2\dot{r}^{4}}{f^{3}}){f'}^2
+(\frac{1}{2}-\frac{\dot{r}^{2}}{2f}-\frac{\dot{r}^{4}}{f^{2}})f^{''}+
\\
&&\hspace{-10mm}
[2\phi'(\rho)+1][\frac{f}{r^2}+\frac{2f'}{r}+\frac{\dot{r}^2}{r^2}+\frac{5f'\dot{r}^2}{2rf}+\frac{\dot{r}^4f'}{rf^2}+\frac{\ell(\ell+1)}{r^2}]
\nonumber
\eea
Note that  in these equations, $\rho$ is the function of $r$. In our numeric results, we will take the angular momentum number $\ell=1$ as illustrations.

\subsection{Stability analysis}

Introducing a new function $\psi$ which is related with $\eta_{\ell m}$ through  definitions (we will focus on fixed $\ell$ and $m$ thus ignore subscripts on the right hand side), 
\bea{}
\eta_{\ell m}=e^{-\int dr(\frac{(2\phi'(\rho)+1)}{2r}-\frac{\rho^{2}r^{2}}{16}\frac{f'}{f^{2}})}\psi
\label{WaveFormPer}
\eea
with the help of \eqref{Kijeq}, the quantity $\eta_{\ell m}$'s equation of motion \eqref{finalPerEQ} can be written into  a Schrodinger-like wave equation
\bea{}
\ddot{\psi} +V_\mathrm{eff}[r(\tau)]\psi =0
,~
V_\mathrm{eff}[r]=-\frac{b^2}{4}-\frac{\dot{b}}{2}+c
\label{WaveEQ}
\eea
where $r(\tau)$ is determined by the unperturbed wormhole throat evolution \eqref{GeMoPoForTSW}. 

Completely parallel with the spherical symmetric perturbation to wormholes at the born-static point $r=r_0$ \eqref{clStabAnalysis}, our general inhomogeneous perturbations to the evolving thin shell also has two fates of evolution. If the effective potential $V_\mathrm{eff}[r(\tau)]$  keeps positive along the joining-shell's evolving process, the quantity $\psi$ will be oscillatory. If at the same time, the transformation factor $\exp\{-\int dr[\frac{2\phi'(\rho)+1}{2r}-\frac{\rho^2r^2}{16}\frac{f'}{f^2}]\}$ does not grow infinitely as time passes by, perturbations $\eta_{\ell m}$ will be stable and the wormhole would be formed through joining two building-block spacetimes at arbitrary initial radius and the following on natural evolutions. Otherwise, perturbations are unstable and the wormhole would be torn up during the following on evolution towards the zero point of $\dot{r}^2\equiv H(r)$. Our numeric results indicate that, for those wormholes cut-and-pasted on the born-static $r_0$ with $H(r_0)=H'(r_0)=0$, $H''(r_0)<0$, both the radial perturbation and general perturbations to the system are stable. However, for those wormholes cut-and-pasted on the non-born-static point with $H(r_0)=H'(r_0)=0$, $0<H''(r_0)$, the perturbation can also be static as long as the system is evolving towards $r_0$ instead of away from it. Let us examine the detailed results below. All our numerics are done with $\ell=1,m=0$, so the lower indices $\ell,m$ in $\eta_{\ell,m}$ will be ignored exclusively.

\section{Numerical results}
\label{Nresults}

\begin{figure}[!ht]
\begin{center}
\includegraphics[scale=0.5]{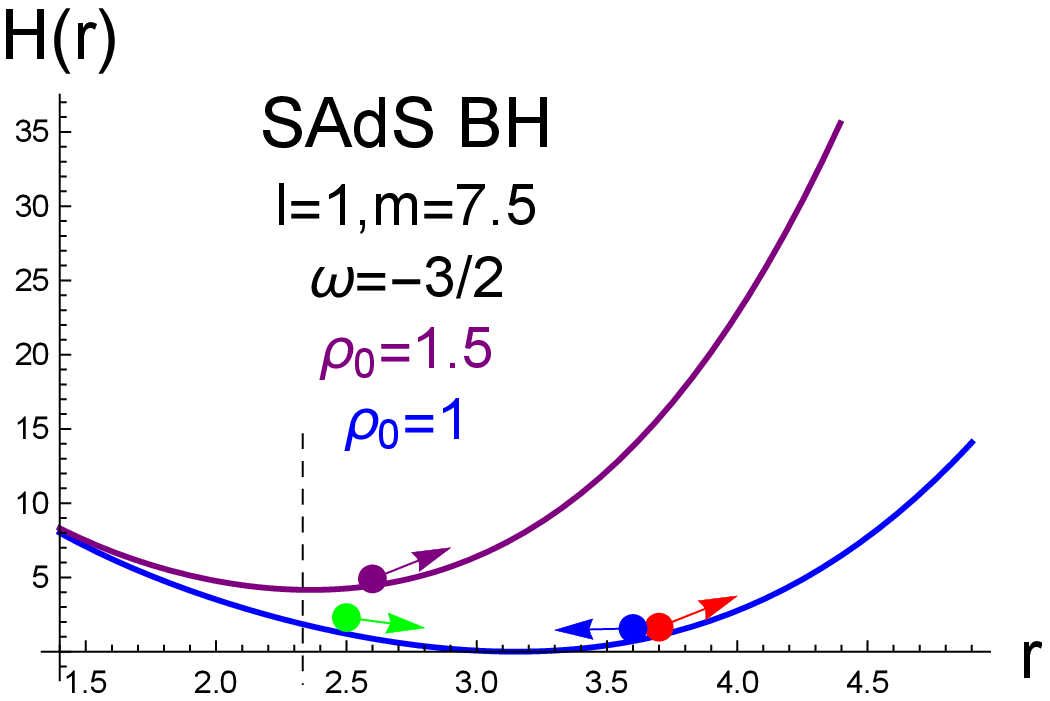}
\includegraphics[scale=0.5]{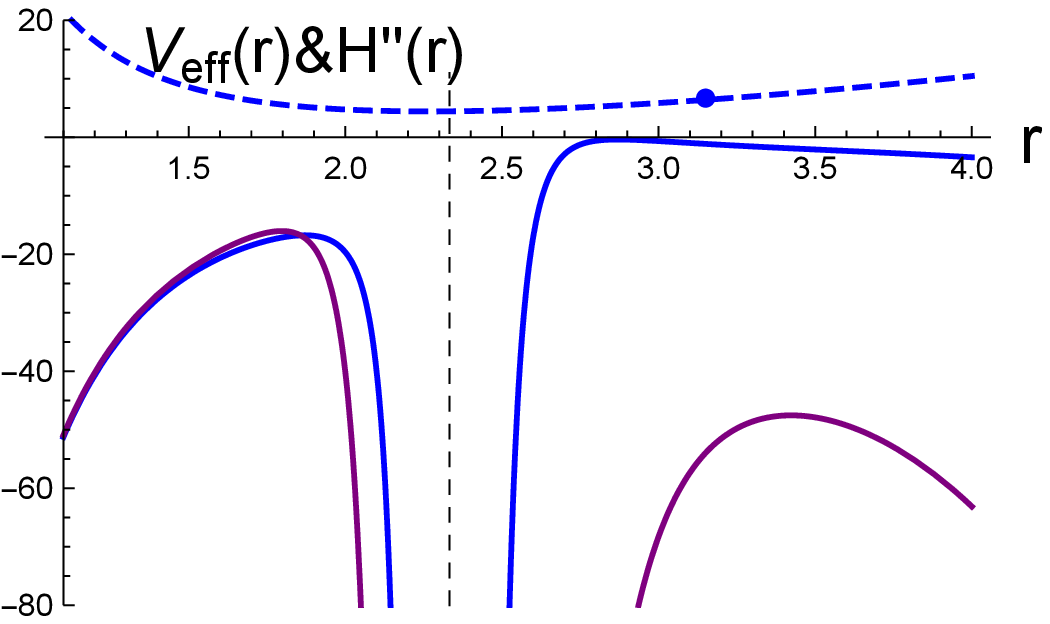}
\includegraphics[scale=0.5]{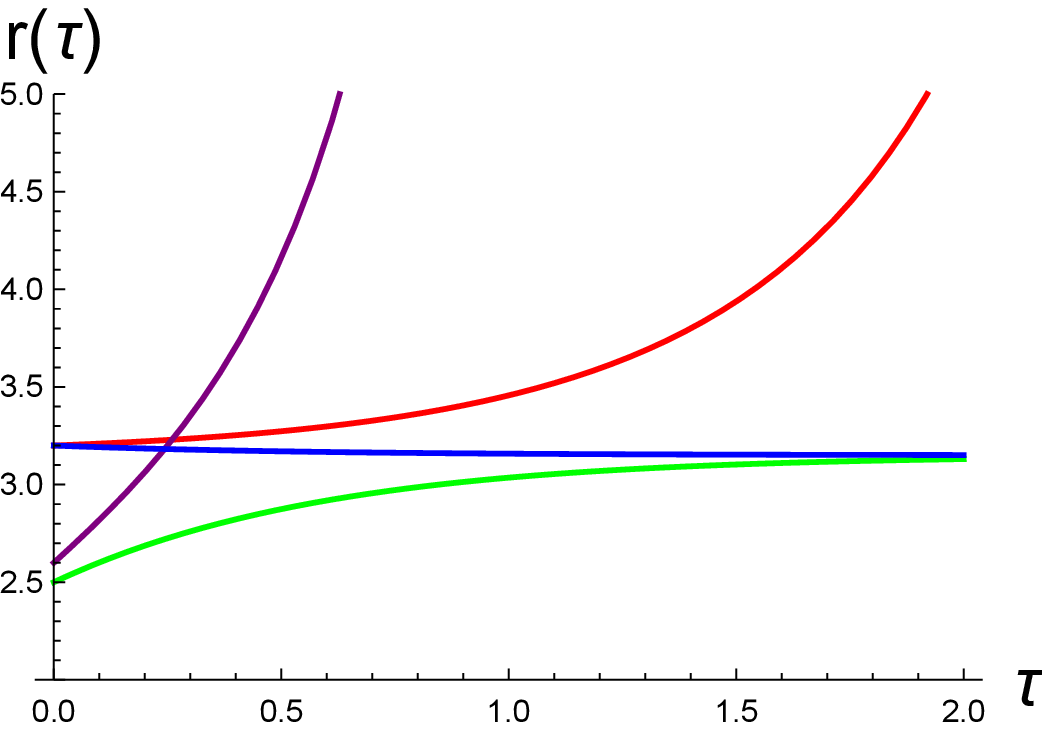}
\includegraphics[scale=0.5]{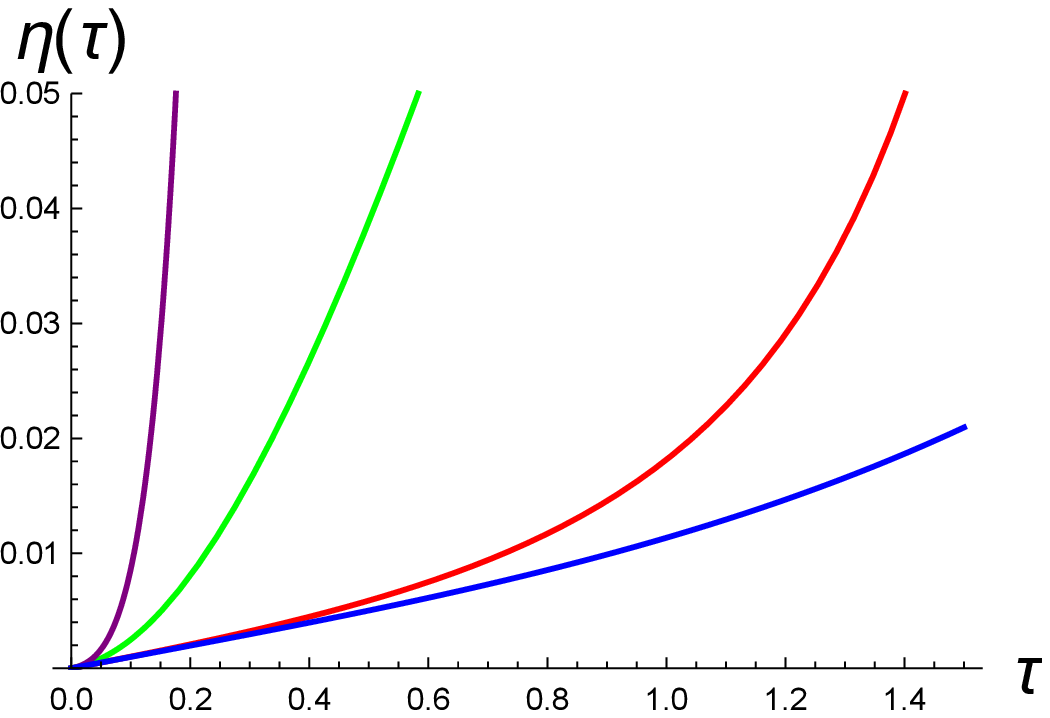}
\caption{(color online) Stability analysis of TSWs constructed by cuting-and-pasting two SAdS blackhole spacetime with the joining shell filled by Phantom Gas. The left hand side displays the joining shell's radius evolution $\dot{r} ^2=H(r)$ and $r(\tau)$. While the right panel displays the evolution of perturbations $\eta=\exp\{-\int dr[\frac{2\phi'(\rho)+1}{2r}-\frac{\rho^2r^2}{16}\frac{f'}{f^2}]\} \cdot \psi(\tau)$ and the effective potential of $\psi$. When $0<H''$ (dashed curve in the top-right sub figure), perturbations to the static wormhole are unstable according to the previous works. While as $V_\mathrm{eff}<0$, perturbations to the wormhole in evolution are unstable according to our analysis in this work. All wormholes in this figure are unstable.}
\label{figLBGSAdS}
\end{center}
\end{figure}

\begin{figure}[!ht]
\begin{center}
\includegraphics[scale=0.5]{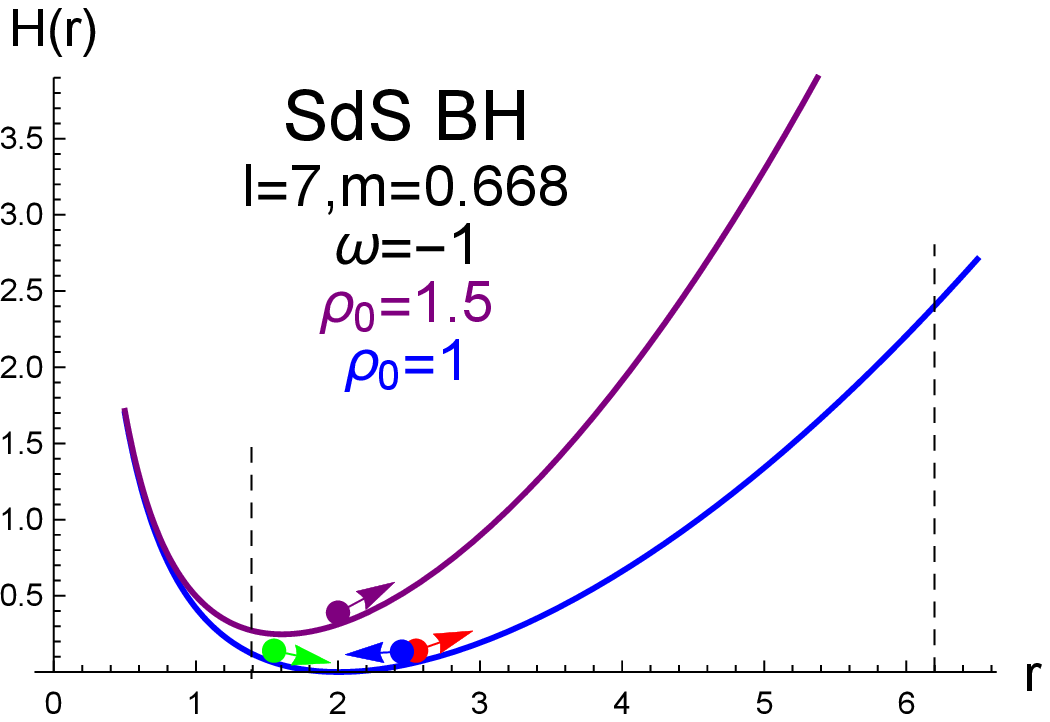}
\includegraphics[scale=0.5]{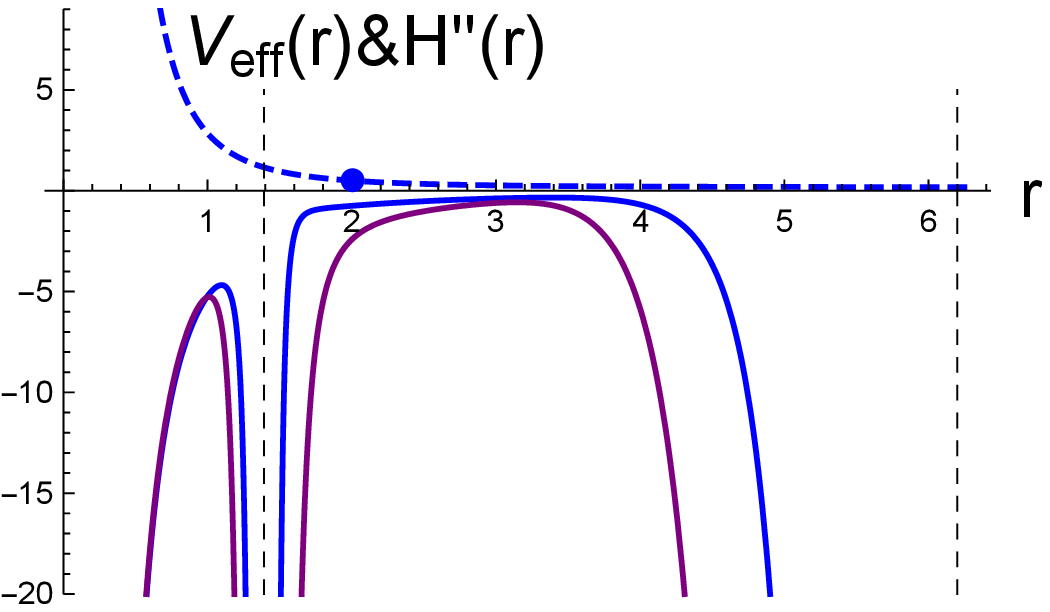}
\includegraphics[scale=0.5]{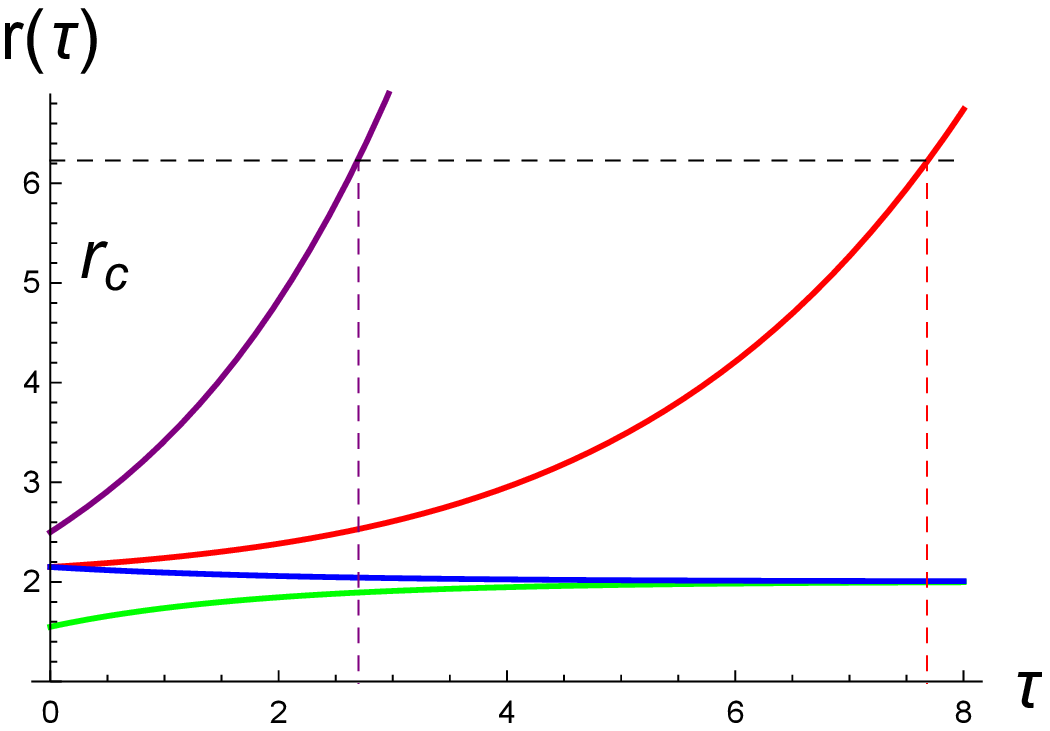}
\includegraphics[scale=0.5]{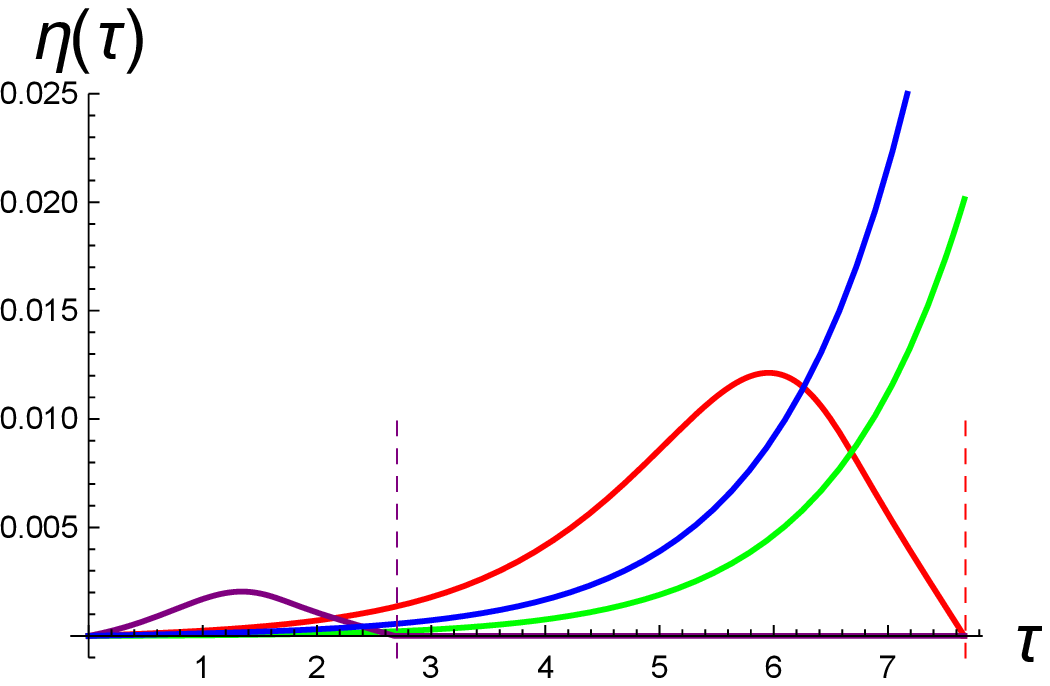}
\caption{(color online) Similar as figure \ref{figLBGSAdS}, but the wormhole is constructed by cutting-and-pasting two SdS blackholes with the joining shell filled by Phantom Gas. The two dashed vertical lines in all subfigures mark the blackhole and cosmic horizon radius (or times needed for the joining shell to evolve to such sizes) of the building-block spacetime.  All wormholes in this figure are unstable}
\label{figLBGSdS}
\end{center}
\end{figure}

\begin{figure}[!ht]
\begin{center}
\includegraphics[scale=0.5]{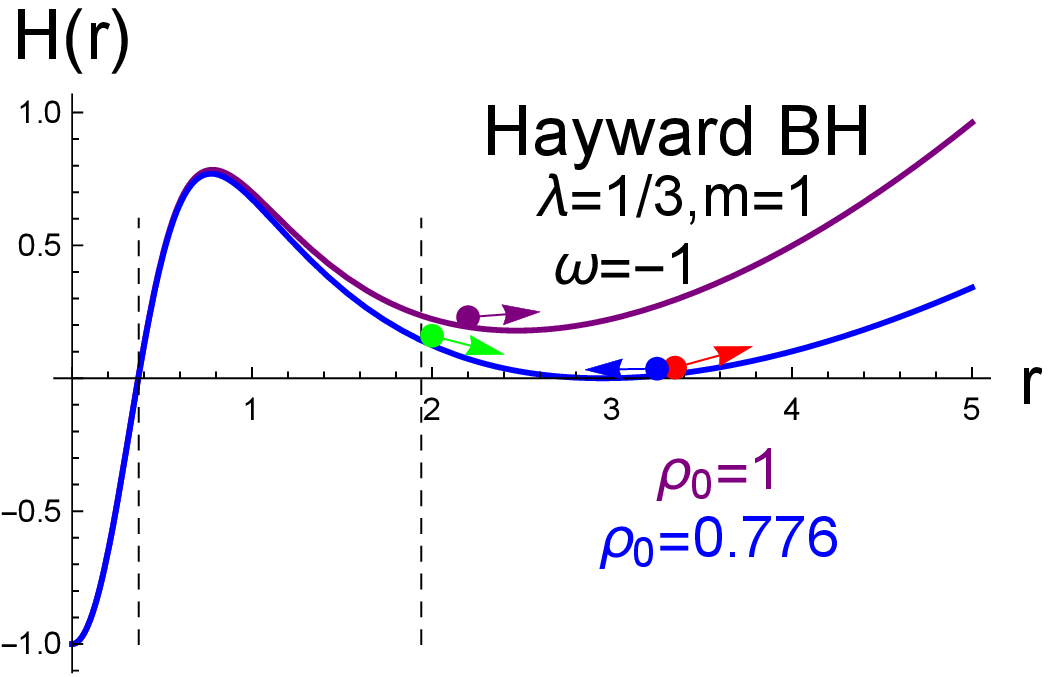}
\includegraphics[scale=0.5]{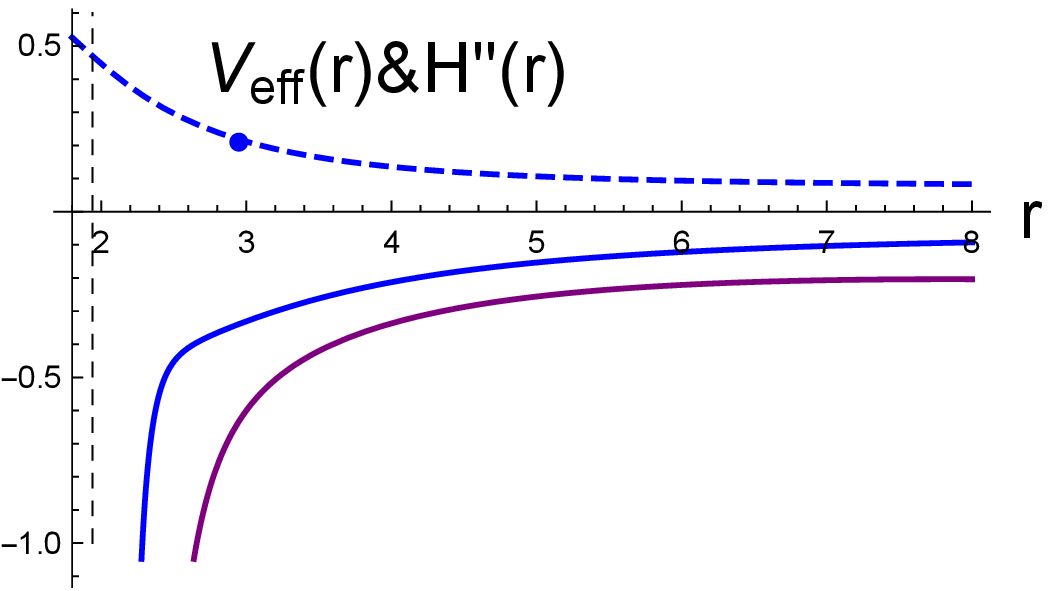}
\includegraphics[scale=0.5]{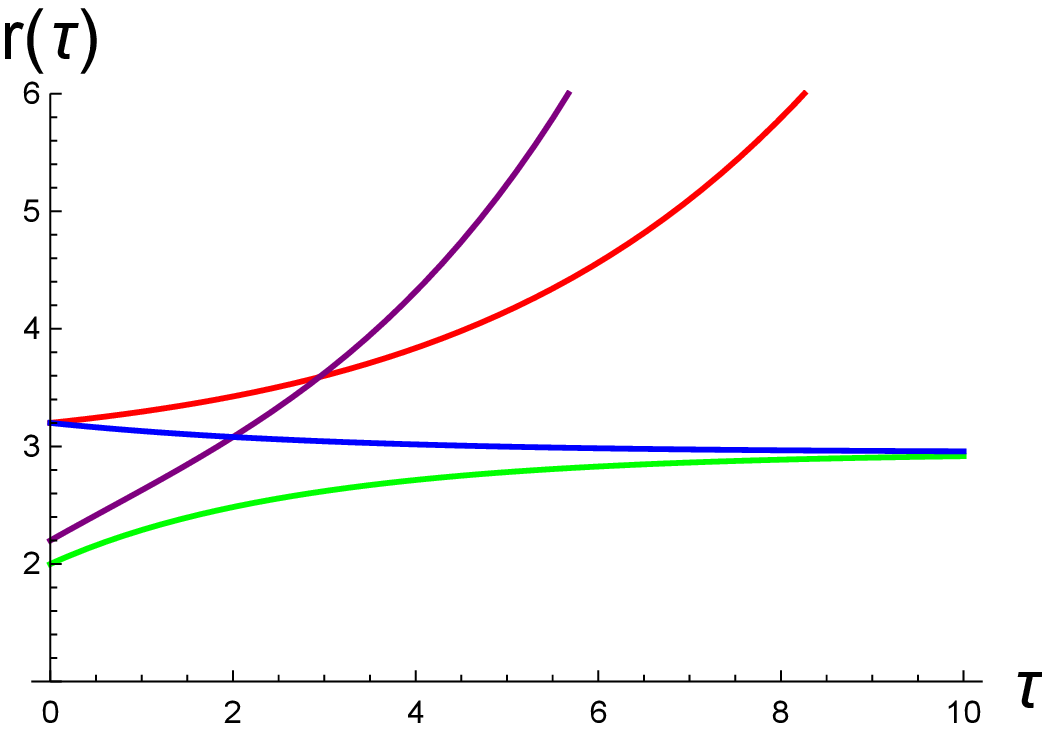}
\includegraphics[scale=0.5]{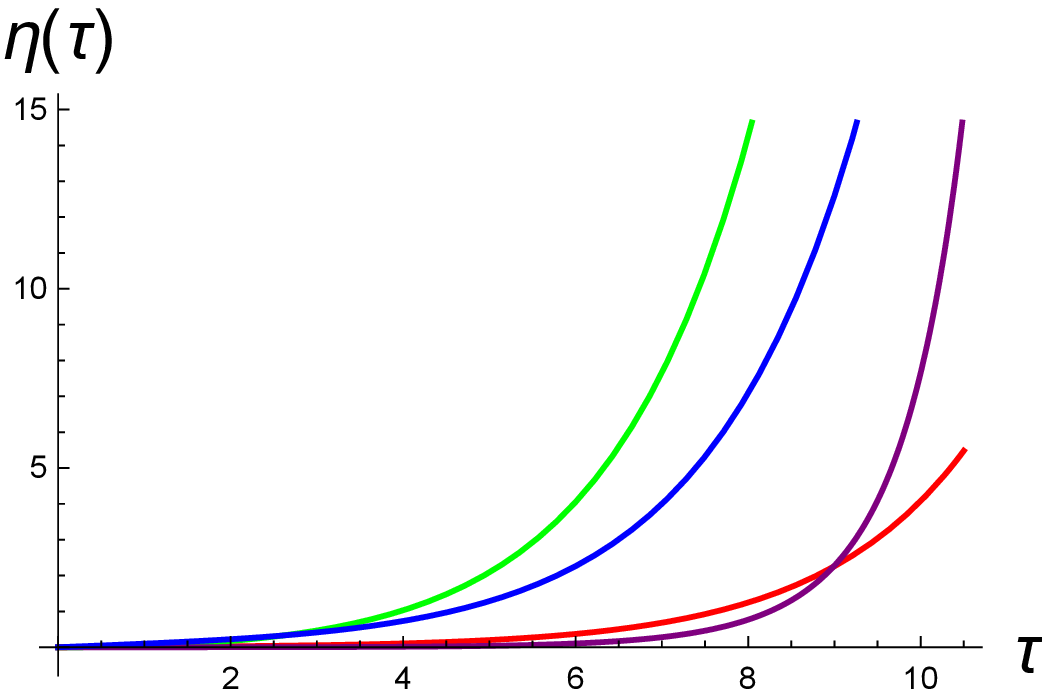}
\caption{(color online) Similar as figure \ref{figLBGSAdS}, but the wormhole is constructed by cutting-and-pasting two Hayward blackholes with the joining shell filled by Phantom Gas. All wormholes in this figure are unstable.}
\label{figLBGHa}
\end{center}
\end{figure}

\begin{figure}[!ht]
\begin{center}
\includegraphics[scale=0.5]{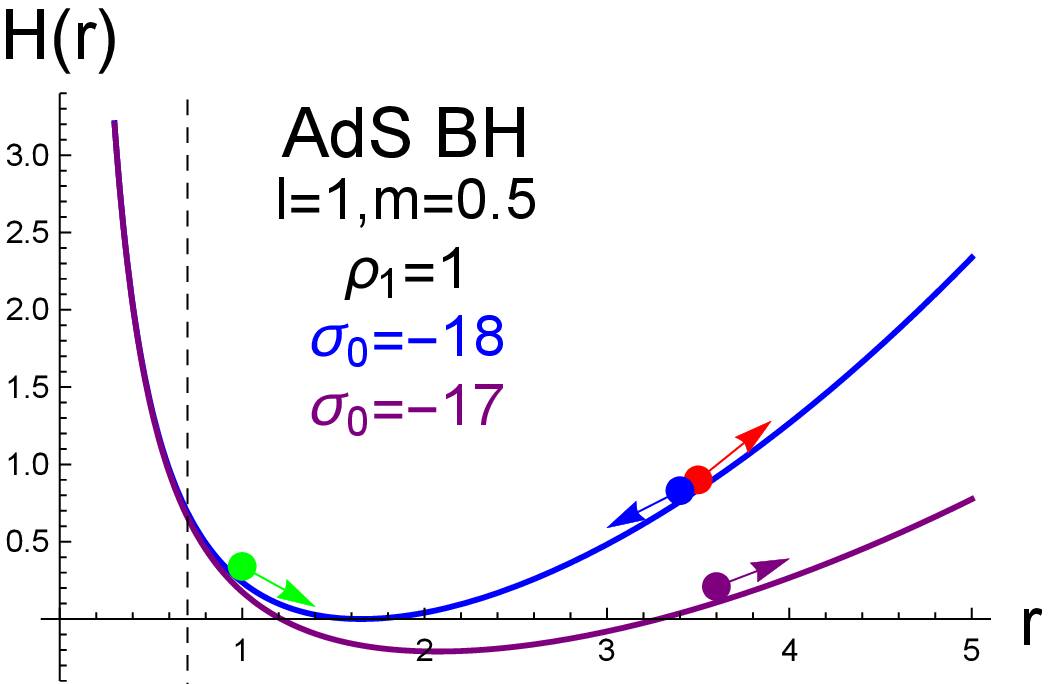}
\includegraphics[scale=0.5]{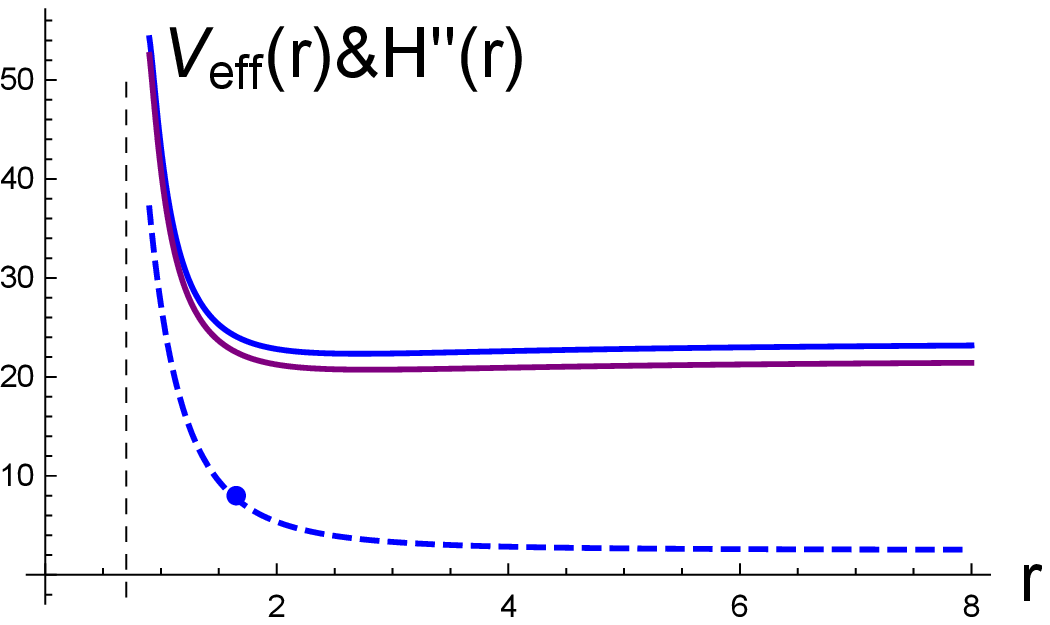}
\includegraphics[scale=0.5]{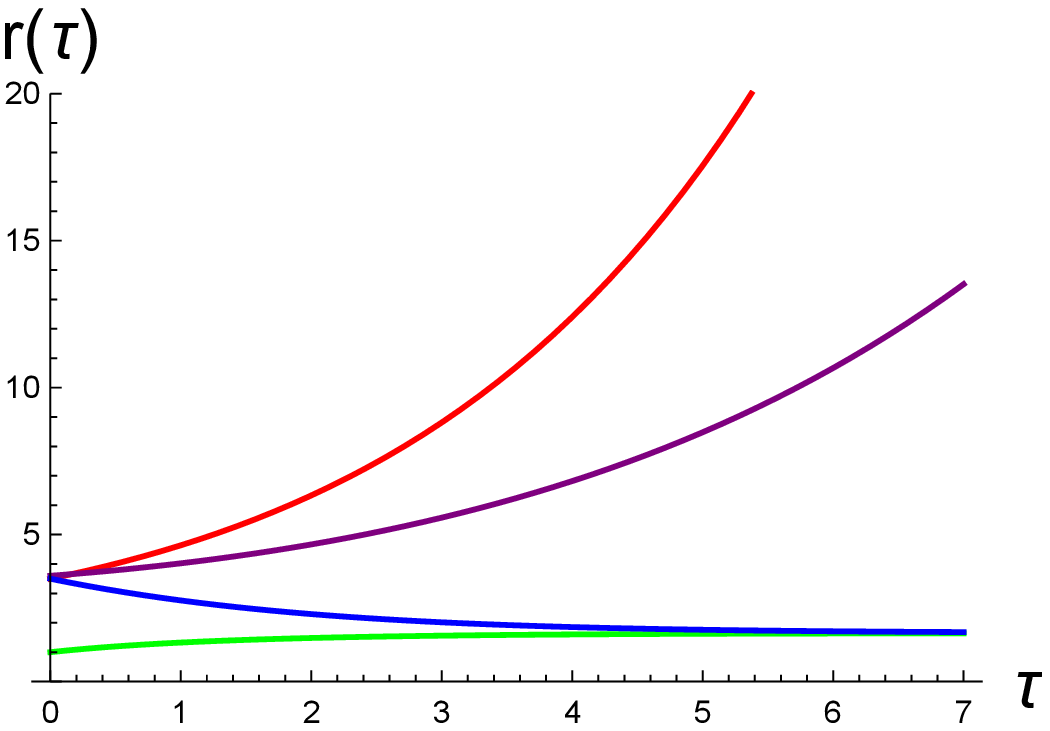}
\includegraphics[scale=0.5]{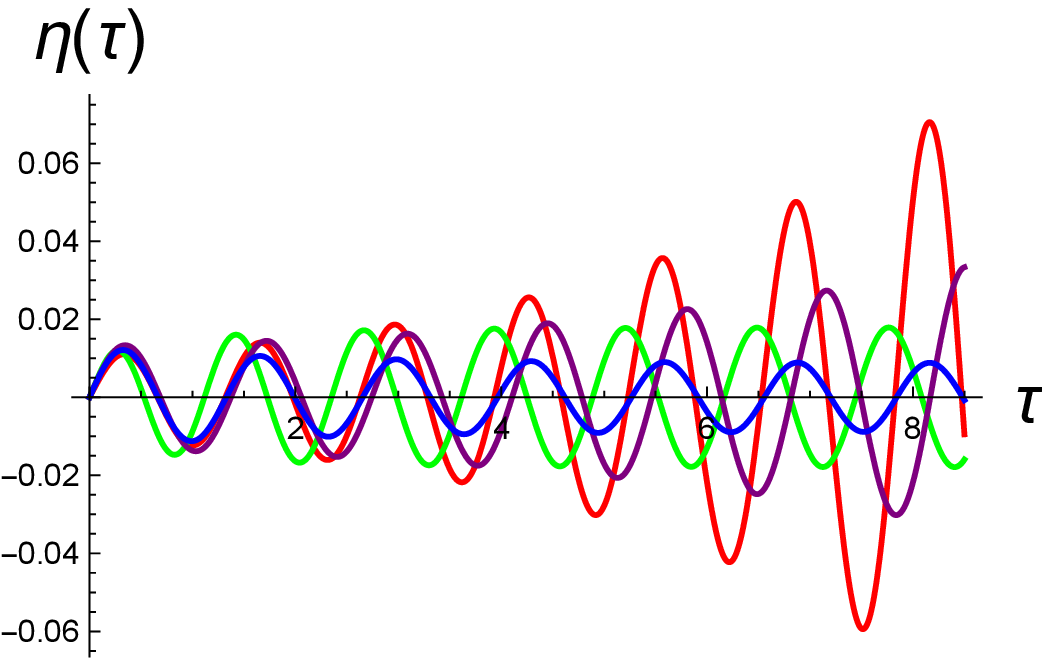}
\caption{(color online) Stability analysis of TSWs constructed by cutting-and-pasting two SAdS blackholes with the joining shell filled by Chaplygin gas (CG). The left hand side displays the throat radius' evolution $\dot{r} ^2=H(r)$ and $r(\tau)$. While the right panel displays the evolution of perturbations $\eta=\exp\{-\int dr[\frac{2\phi'(\rho)+1}{2r}-\frac{\rho^2r^2}{16}\frac{f'}{f^2}]\}\cdot \psi(\tau)$ and the effective potential of $\psi$. As $0<V_\mathrm{eff}$, a wormhole initially constructed by cut-and-paste building-block spacetimes at non-$r_0$ radius thin shell can be stable to perturbations if it evolves towards $r_0$, see the green and blue curves in the bottom right figure. If the wormhole evolves away from $r_0$, it is unstable under perturbations, see the red and purple curve in the bottom right figure.}
\label{figCGSAdS}
\end{center}
\end{figure}

\begin{figure}[!ht]
\begin{center}
\includegraphics[scale=0.5]{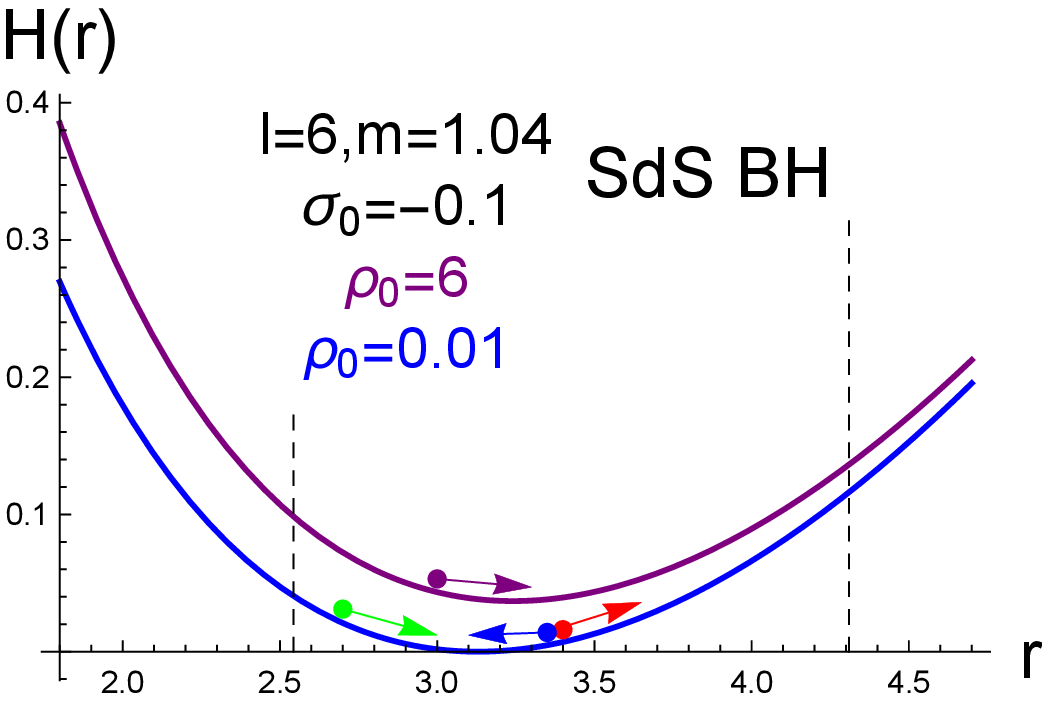}
\includegraphics[scale=0.5]{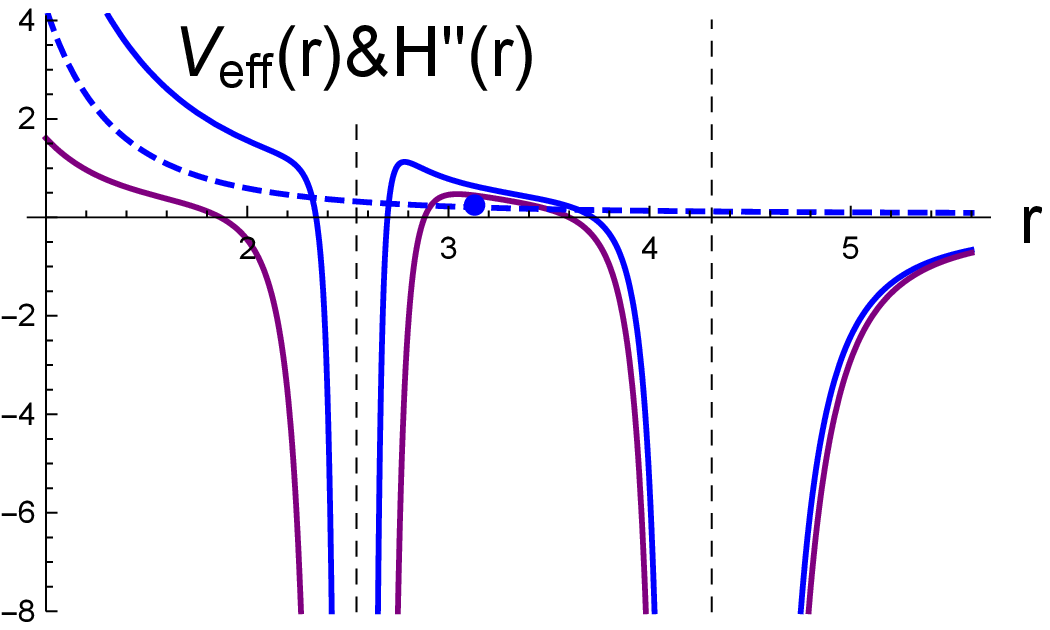}
\includegraphics[scale=0.5]{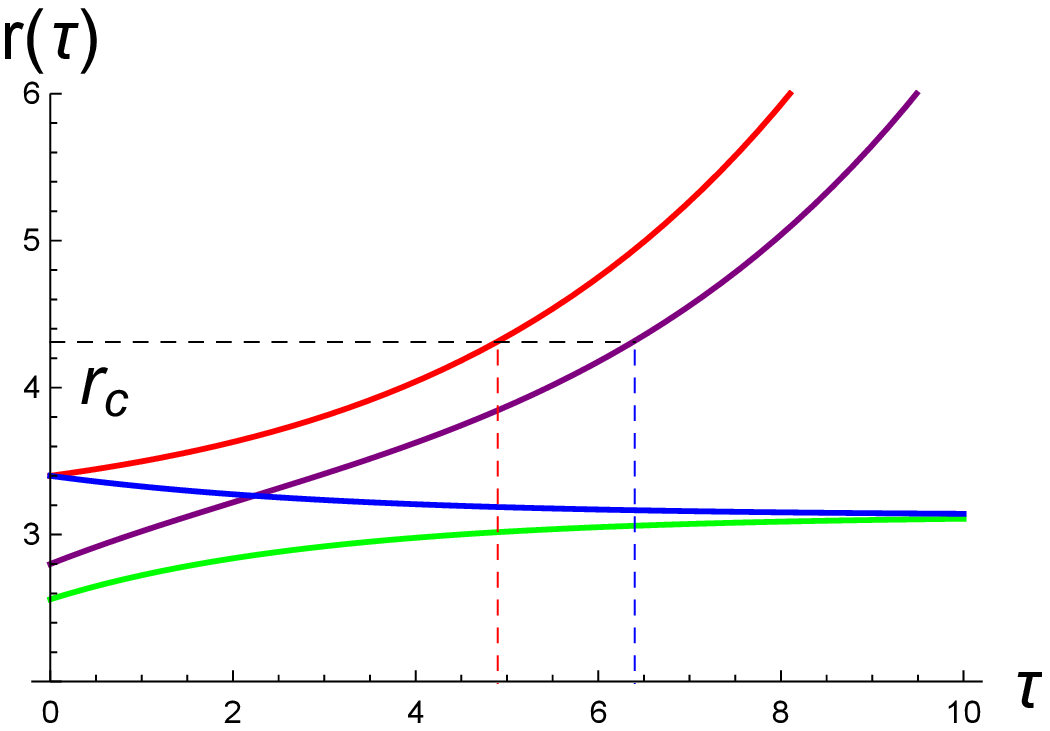}
\includegraphics[scale=0.5]{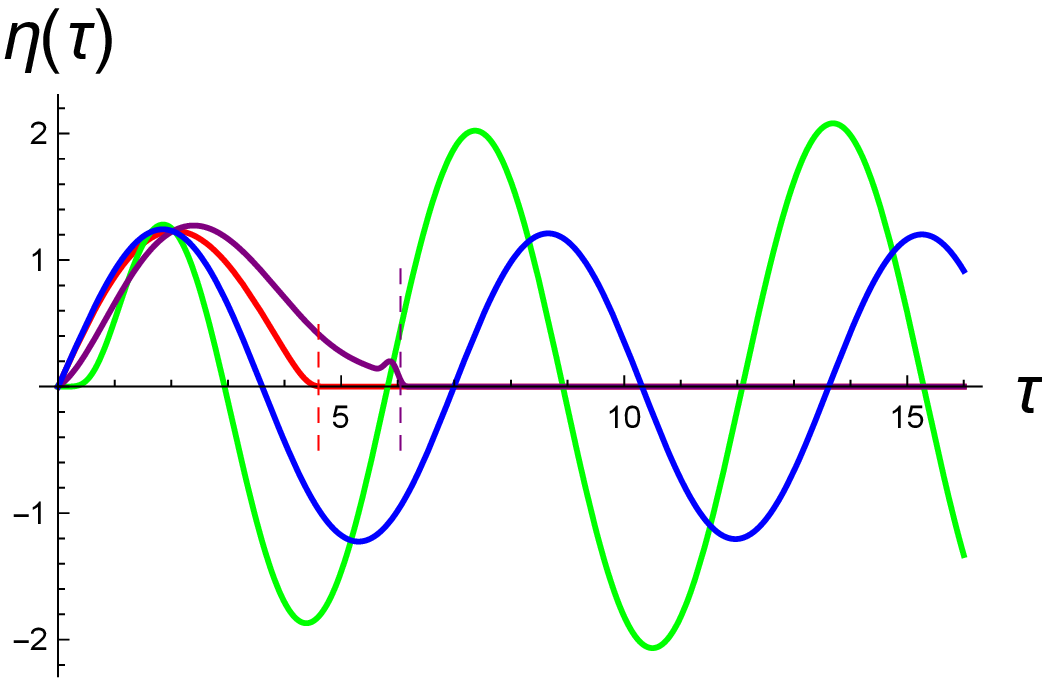}
\caption{(color online)Similar as figure \ref{figCGSAdS}, but the wormhole is constructed by cut-and-paste two SdS blackhole spacetimes. All perturbations to wormholes in this figure are do not diverge in the sub-cosmic-horizon region. However, no static wormhole exists for parameter choice in this figure.}
\label{figCGSdS}
\end{center}
\end{figure}

\begin{figure}[!ht]
\begin{center}
\includegraphics[scale=0.5]{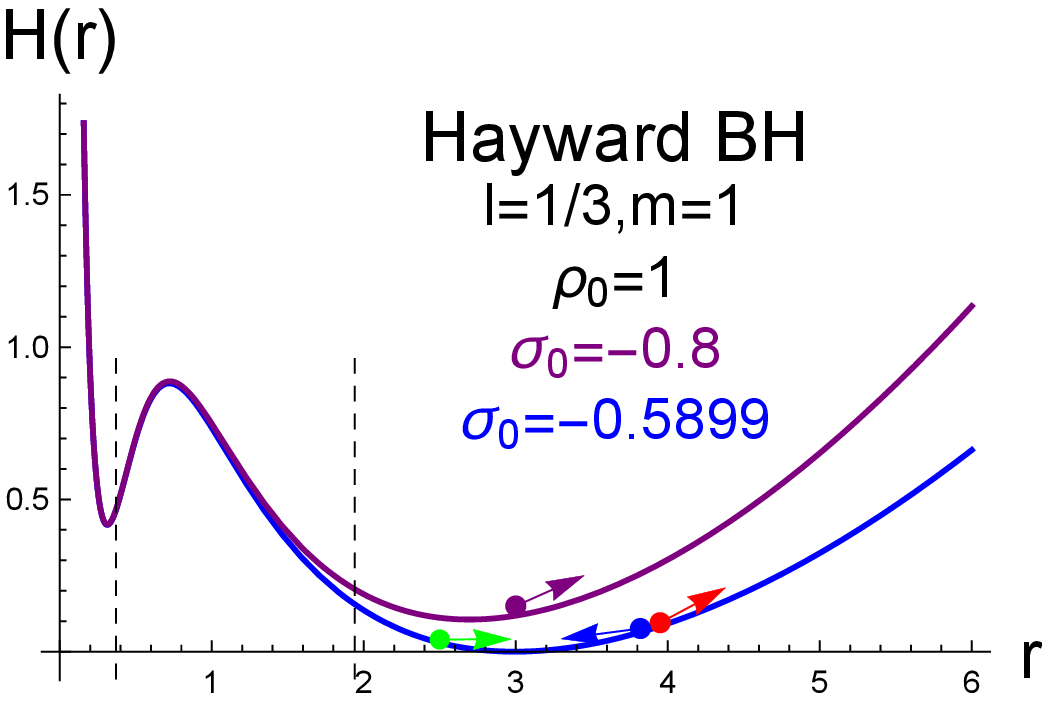}
\includegraphics[scale=0.5]{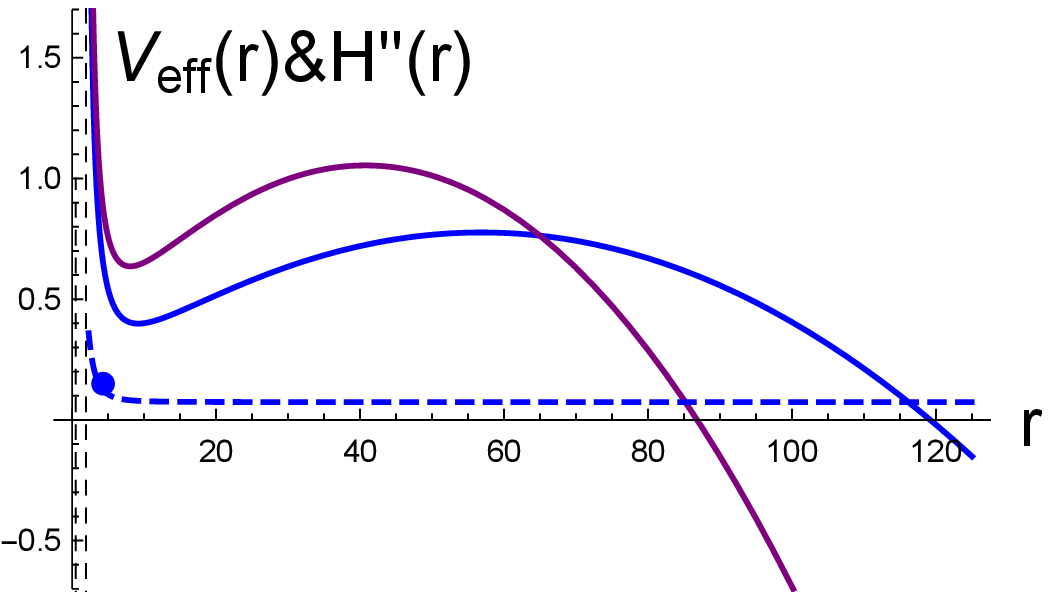}
\includegraphics[scale=0.5]{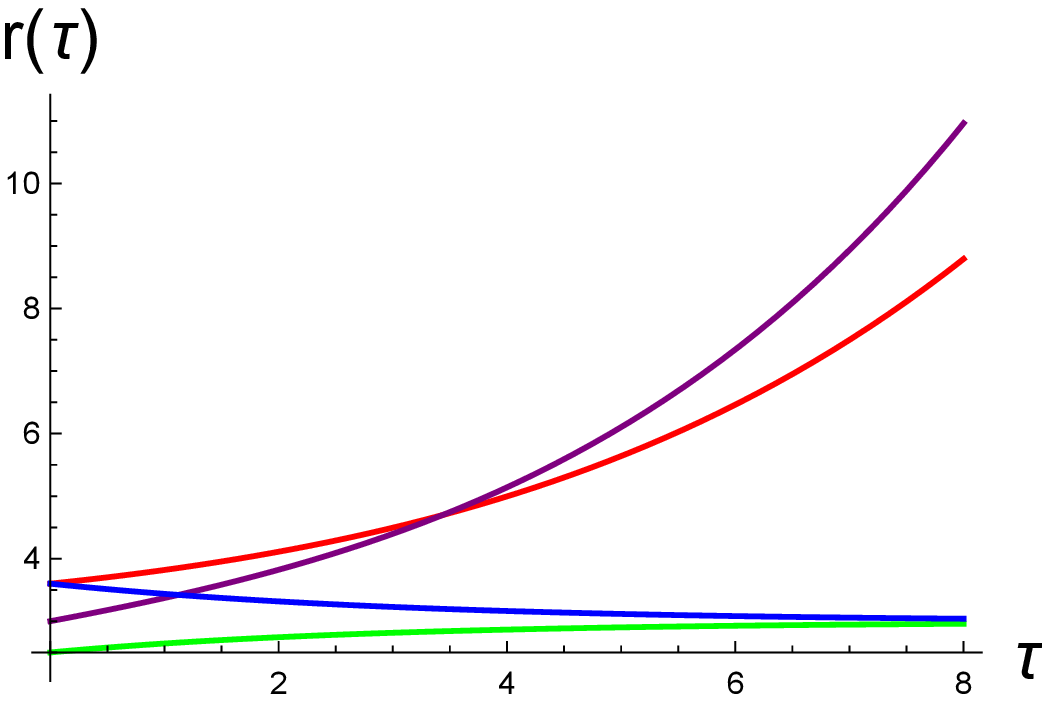}
\includegraphics[scale=0.5]{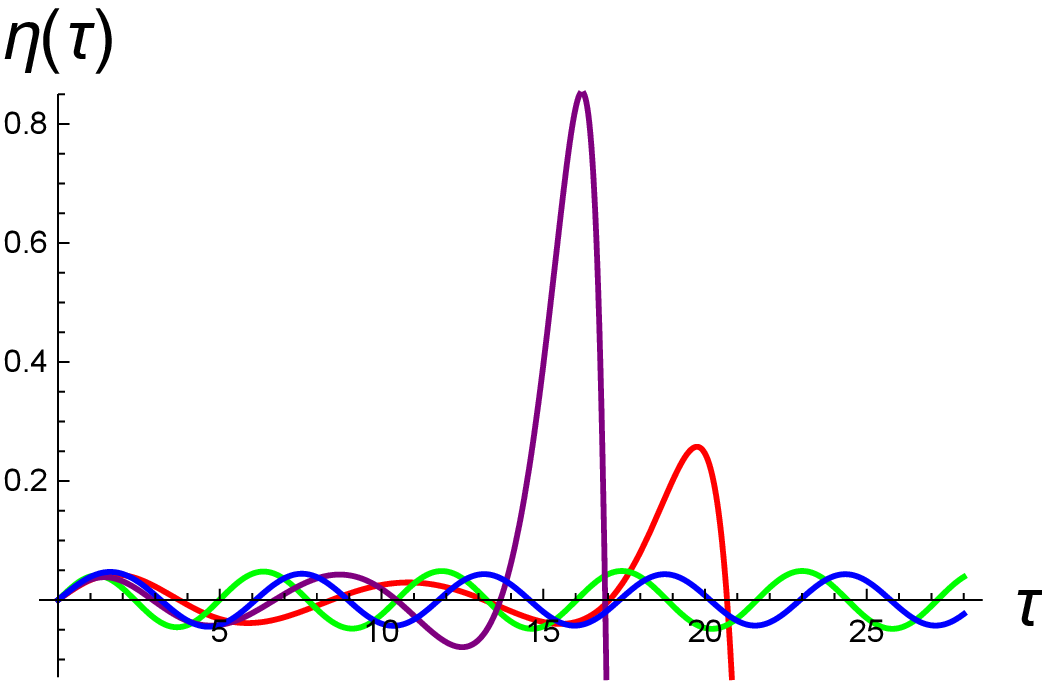}
\caption{(color online) Similar as figure \ref{figCGSAdS}, but the wormhole is constructed with two Hayward blackhole spacetime with the throat filled by CG. As long as the initial direction is towards $r_0$, the wormhole cut-and-paste at non-$r_0$ joining-shell is stable to perturbation and will evolve to $r_0$ naturally.}
\label{figCGHa}
\end{center}
\end{figure}

Numerically we need to solve the wormhole joining shell's classical evolution equation $\dot{r}^2\equiv H(r)=\rho^2(r)r^2/16-f(r)$ for given parameter choices, and substitute the resultant $r(\tau)$ into the perturbation equation \eqref{WaveEQ} to see $\psi(\tau)$'s evolution. Of course, the final conclusion on stability should be made according to the behavior of perturbative quantity $\eta(\tau)$ defined in eq.\eqref{WaveFormPer}. Our results are displayed in figure \ref{figLBGSAdS}-\ref{figCGHa} collectively. In figure \ref{figLBGSAdS}-\ref{figLBGHa}, the matters filling in the wormhole throat are Phantom Gas featured by equation of state $P=\omega\rho$ with $\omega\leqslant-1$($w=-1$ corresponds with the usual dark energy of cosmological constant), while in figure \ref{figCGSAdS}-\ref{figCGHa}, the filling matters are the Chaplygin gas featured by exotic equation of state $P\cdot\rho=\sigma_0$. 

Figure \ref{figLBGSAdS}-\ref{figLBGHa} displays the stability analysis for three types of wormhole constructed from cutting-and-pasting two copies of Schwarzschild-Anti-deSitter, Schwarzschild-deSitter and Hayward blackholes respectively, with the joining shell filled with Phantom Gas. For each type of wormholes, we consider two kinds of joining shells' evolving trend. The first is towards a final extremal point $r_0$ defined by $\dot{r}_0\propto H(r_0)=0$, $\ddot{r}_0\propto H'(r_0)=0$ and plotted with green and blue curves in the figure. From the figure we easily see that, in all three building-block spacetimes, the wormholes are all unstable to perturbations. The second is away from the extremal point and plotted with red and purple curves.  From the figure we see that in both SAdS and Hayward black-hole building-block spacetimes, the perturbations are unstable. However, in the SdS building-blocks, the perturbations do not diverge when the joining shell radius grow and approach the cosmic horizon size. Nevertheless, wormholes joined at such joining shells cannot be static, because $\dot{r}$ at such values of the shell radius $r=r_c$ is nonzero. So general lessons following from these figures are, although in all three types of wormholes, extremal point of the joining shell radius $r_0$ exists for which $\dot{r}_0\propto H(r_0)=0$, $\ddot{r}_0\propto H'(r_0)=0$. Wormholes constructed by cutting-and-pasting building-blocks at such radius are unstable, both under spherical symmetric perturbations to the system exactly joined at such point and general perturbations to the system joined away from but be experiencing evolutions to such point. Of course, since we make no complete scanning of the full  parameter space $\omega$-$\rho_0$, we cannot say that phantom gases cannot support stable wormholes in all three building-block spacetimes.



Figure \ref{figCGSAdS}-\ref{figCGHa} displays the stability analysis for three types of wormhole constructed from cutting-and-pasting two copies of Schwarzschild-Anti-deSitter, Schwarzschild-deSitter and Hayward-Minkowski blackholes respectively, with the joining shell filled with Chaplygin Gas. Just the same as PG supported wormholes discussed in the previous paragraph,  perturbations to the CG wormholes will also occur in two directions in this paragraph. In the first case, the joining shell is evolving towards a final extremal point and the corresponding quantities are plotted with green and blue curves in the figure, while in the second case, the joining shell is evolving away from such point and the corresponding quantities are plotted with red and purple curves in the figure. From the figure, we easily see in the first case, perturbations in all three building-block spacetimes have all constant amplitude oscillatory feature as time passes by. As results, wormholes in this case can be joined at an arbitrary initial radius and experience a dynamic evolving process, and arrive on a final stable point naturally. While in the second case, the wormholes are all unstable under perturbations, the joining shell will be torn up during the evolution away from $r_0$. 

Comparing analysis in figure \ref{figLBGSAdS}-\ref{figLBGHa} and  \ref{figCGSAdS}-\ref{figCGHa}, we can conclude  that, the stability of a wormhole under general linear perturbations are mainly determined by two factors, the first is the equation of state of matters filled in the joining shell, the second is the evolving direction of joining shells at initial times. CG-like filling matter and extremal-point towards joining shell radius' evolution lead to stable wormholes. In other cases, the wormhole are unstable. 

Comparing our results in this work and those in literatures \cite{Poisson:1995sv, Lobo:2003xd, Lobo:2005yv, Eiroa:2007qz}, we see that, \{$H(r_0)=0$, $H'(r_0)=0$, $H''(r_0)<0$\} are non necessary conditions for getting stable wormholes by the cut-and-paste method. In the case \{$H(r_0)=0$, $H'(r_0)=0$, $0<H''(r_0)$\}, stable wormholes are also possible when the joining-shell radius is initially unequal to $r_0$ but is evolving towards $r_0$ instead of away from such values. That is, we can cut-and-paste two building-block spacetimes at $r$-values of the joining-shell radius and let the system evolving dynamically towards $r_0$, the system will keep stable to general perturbations during the whole evolving process until the final $r_0$-value is arrived on.

\section{Conclusion and Discussion}
By the standard cut-and-paste method, we construct spherical thin-shell wormholes supported by Phantom-gas and Chaplygin-gas with three kinds of building-block spacetimes, the Schwarzchild-AdS, Schwarzchild-dS and Haward non-singular blackholes. Borrowing ideas from the cosmological perturbation theory, we consider scalar type inhomogeneous and anisotropic perturbations $\zeta(x)$ to the evolving thin-shell joining the two building-block spacetimes, with the source identified as the quantum fluctuation of matters living on the wormhole throat, i.e. the joining-shell. We derive out equations of motion for $\zeta$ and its polar coordinate decomposition $\zeta (\tau,\theta,\varphi)=\sum_{\ell m}\eta_{\ell m}(\tau)Y_{\ell m}(\theta,\varphi)$. By numeric methods, we compute time evolutions of $\eta_{\ell m}(\tau)$ and make stability analysis accordingly. Comparing with previous works in this area, we for the first time, make this analysis on an evolving wormhole joining shell, while all previous works execute their analysis on fixed and born static joining-shells.

We consider two kinds of filling matter, i.e. Phantom-gas and Chaplygin-gas in the joining-shell, each were used in three kinds of building-block spacetimes, i.e. Schwarzschild-Anti-deSitter, Schwarzschild-deSitter and Hayward blackholes. Our numerical results indicate that, for the first filling matter, wormholes constructed by cutting-and-pasting all three building-block spacetimes are unstable. While for the second filling matter, stable wormholes can be obtained from this cut-and-paste method. Especially, we find that  \{$H(r_0)=H'(r_0)=0$, $H''(r_0)<0$\} are not necessary conditions for getting stable wormholes. In the case \{$H(r_0)=H'(r_0)=0$, $0<H''(r_0)$\}, stable wormholes are also possible when the joining-shell radius is initially unequal to $r_0$ but is evolving towards $r_0$ instead of away from such values. This forms contrast with conclusions in the literature but is a natural way the wormhole evolves to the static point when it is constructed at a non-born-static point. 

As discussions, we note here that two urgent questions need be explored in the near future. The first is, will the statement that \{$H(r_0)=H'(r_0)=0$, $H''(r_0)<0$\} are non-necessary condition for stable wormholes still be true when higher order perturbations are considered? Since in second order perturbations, $r^{(0)}_\mathrm{bg}(\tau)+\delta^{(1)} r(\tau)$ becomes the new background $r_\mathrm{bg}$, whose evolving direction will be highly dependent on the first order perturbation $\delta^{(1)}r(\tau)$ and will be non-uniformly towards $r_0$ any more. In such cases, we have to worry about the system's ability to keep stable for further.  The second is, what's the effect of bulk space's perturbation to the stability of thin shell wormholes? Obviously in the current work, we considered only perturbations of the thin shell wormhole themselves. Other questions like the perturbation to dynamically evolving wormholes in higher derivative gravitations, such as f(R), Gauss-Bonnet, Infinite derivatives gravitation, etc are also interesting  for future works.

\section*{Acknowledgements}
This work is supported by NSFC grant no. 11875082.


\begin{thebibliography}{99}

    \bibitem{HGEllis1973}
    H. G. Ellis, 
    ``Ether flow through a drainhole:  A particle model in general relativity''
    {\em Acta Phys. Pol. B} {\bf 4} (1973) 251.

    \bibitem{Visser1995}
     M. Visser,
    ``Lorentzian Wormholes: From Einstein to Hawking''
    (AIP, New York, 1995).
		
	\bibitem{Morris:1988cz} 
	M.~S.~Morris and K.~S.~Thorne,
	``Wormholes in space-time and their use for interstellar travel: A tool for teaching general relativity,''
	{\it Am.\ J.\ Phys.\ }{\bf 56} (1988) 395.
	
	\bibitem{Hochberg:1997wp}
	D.~Hochberg and M.~Visser,
	``Geometric structure of the generic static traversable wormhole throat,''
	{\it Phys.\ Rev.\ D} {\bf 56} (1997) 4745
	\href{https://arxiv.org/abs/gr-qc/9704082}{arXiv: gr-qc/9704082}	
	
	\bibitem{Morris:1988tu}  
	M.~S.~Morris, K.~S.~Thorne and U.~Yurtsever,
	``Wormholes, Time Machines, and the Weak Energy Condition,''
	{\it Phys.\ Rev.\ Lett.\ } {\bf 61} (1988) 1446.
	
	\bibitem{Hochberg:1998ii}
	D.~Hochberg and M.~Visser,
	``The Null energy condition in dynamic wormholes,''
	{\it Phys.\ Rev.\ Lett.\ } {\bf 81} (1998) 746
	\href{https://arxiv.org/abs/gr-qc/9802048}{arXiv: gr-qc/9802048}
	
	\bibitem{Hochberg:1998vm}
	D.~Hochberg, C.~Molina-Paris and M.~Visser,
	``Tolman wormholes violate the strong energy condition,''
	{\it Phys.\ Rev.\ D } {\bf 59} (1999) 044011
	\href{https://arxiv.org/abs/gr-qc/9810029}{arXiv: gr-qc/9810029}
		
	\bibitem{Visser:1989kh}
	M.~Visser,
	``Traversable wormholes: Some simple examples,''
	{\it Phys.\ Rev.\ D} {\bf 39} (1989) 3182
	\href{https://arxiv.org/abs/0809.0907}{arXiv:0809.0907[gr-qc]}
	
	\bibitem{Visser:1989kg} 
	M.~Visser,
	``Traversable wormholes from surgically modified Schwarzschild space-times,''
	{\it Nucl.\ Phys.\ B} {\bf 328} (1989) 203
	\href{https://arxiv.org/abs/0809.0927}{arXiv:0809.0927 [gr-qc]}.
	
	\bibitem{Bejarano:2006uj}
	C.~Bejarano, E.~F.~Eiroa and C.~Simeone,
	``Thin-shell wormholes associated with global cosmic strings,''
	{\it Phys.\ Rev.\ D} {\bf 75} (2007) 027501
	\href{https://arxiv.org/abs/gr-qc/0610123}{arXiv: gr-qc/0610123}
		
	\bibitem{Bhawal:1992sz}
	B.~Bhawal and S.~Kar,
	``Lorentzian wormholes in Einstein-Gauss-Bonnet theory,''
	{\it Phys.\ Rev.\ D} {\bf 46} (1992) 2464.
	doi:10.1103/PhysRevD.46.2464
	
	\bibitem{Hochberg:1990is}
	D.~Hochberg,
	``Lorentzian wormholes in higher order gravity theories,''
	{\it Phys.\ Lett.\ B} {\bf 251} (1990) 349.
	doi:10.1016/0370-2693(90)90718-L
	
	\bibitem{Agnese:1995kd} 
	A.~G.~Agnese and M.~La Camera,
	``Wormholes in the Brans-Dicke theory of gravitation,''
	{\it Phys.\ Rev.\ D} {\bf 51} (1995) 2011.
	doi:10.1103/PhysRevD.51.2011
	
	\bibitem{Eiroa:2005pc} 
	E.~F.~Eiroa and C.~Simeone,
	``Thin-shell wormholes in dilaton gravity,''
	{\it Phys.\ Rev.\ D} {\bf 71} (2005) 127501
	\href{https://arxiv.org/abs/gr-qc/0502073}{arXiv: gr-qc/0502073}
	
	\bibitem{Lemos:2004vs} 
	J.~P.~S.~Lemos and F.~S.~N.~Lobo,
	``Plane symmetric traversable wormholes in an Anti-de Sitter background,''
	{\it Phys.\ Rev.\ D} {\bf 69} (2004) 104007
	\href{https://arxiv.org/abs/gr-qc/0402099}{arXiv:gr-qc/0402099}
		
	\bibitem{Clement:1997yp}
	G.~Clement,
	``Flat wormholes from straight cosmic strings,''
	{\it J.\ Math.\ Phys.\ } {\bf 38} (1997) 5807
	\href{https://arxiv.org/abs/gr-qc/9701060}{arXiv: gr-qc/9701060}

	\bibitem{Lobo:2005us}
     F.~S.~N.~Lobo,
    ``Phantom energy traversable wormholes,''
     {\it Phys.\ Rev.\ D} {\bf 71} (2005) 084011
     \href{https://arxiv.org/abs/gr-qc/0502099}{arXiv: gr-qc/0502099}

	\bibitem{Lobo:2005vc} 
	F.~S.~N.~Lobo,
	``Chaplygin traversable wormholes,''
	{\it Phys.\ Rev.\ D} {\bf 73}, 064028 (2006)
	\href{http://arxiv.org/abs/gr-qc/0511003}{arXiv:0511003 [gr-qc]}
		
	\bibitem{Poisson:1995sv} 
	E.~Poisson and M.~Visser,
	``Thin shell wormholes: Linearization stability,''
	{\it Phys.\ Rev.\ D} {\bf 52} (1995) 7318
	\href{https://arxiv.org/abs/gr-qc/9506083}{arXiv:gr-qc/9506083}
	
	\bibitem{Lobo:2003xd} 
	F.~S.~N.~Lobo and P.~Crawford,
	``Linearized stability analysis of thin shell wormholes with a cosmological constant,''
	{\it Class.\ Quant.\ Grav.\ }{\bf 21} (2004) 391
	\href{https://arxiv.org/abs/gr-qc/0311002}{arXiv:gr-qc/0311002}
		
	\bibitem{Lemos:2008aj}
	J.~P.~S.~Lemos and F.~S.~N.~Lobo,
	``Plane symmetric thin-shell wormholes: Solutions and stability,''
	{\it Phys.\ Rev.\ D} {\bf 78} (2008) 044030
	\href{https://arxiv.org/abs/0806.4459}{arXiv:0806.4459 [gr-qc]}
	
	\bibitem{Sharif:2013efa} 
	M.~Sharif and M.~Azam,
	``Mechanical Stability of Cylindrical Thin-Shell Wormholes,''
	{\it Eur.\ Phys.\ J.\ C} {\bf 73} (2013) no.4, 2407
	\href{https://arxiv.org/abs/1308.0196}{arXiv:1308.0196 [gr-qc]}
	
	\bibitem{Garattini:2008xz} 
	R.~Garattini and F.~S.~N.~Lobo,
	``Self-sustained traversable wormholes in noncommutative geometry,''
	{\it Phys.\ Lett.\ B} {\bf 671} (2009) 146
	\href{https://arxiv.org/abs/0811.0919}{arXiv:0811.0919 [gr-qc]}
	
	\bibitem{Lobo:2009ip}
	F.~S.~N.~Lobo and M.~A.~Oliveira,
	``Wormhole geometries in f(R) modified theories of gravity,''
	{\it Phys.\ Rev.\ D} {\bf 80} (2009) 104012
	\href{https://arxiv.org/abs/0909.5539}{arXiv:0909.5539 [gr-qc]}
	
	\bibitem{Garcia:2010xb}
	N.~M.~Garcia and F.~S.~N.~Lobo,
	``Wormhole geometries supported by a nonminimal curvature-matter coupling,''
	{\it Phys.\ Rev.\ D} {\bf 82} (2010) 104018
	\href{https://arxiv.org/abs/1007.3040}{arXiv:1007.3040 [gr-qc]}

	\bibitem{Lobo:2005yv} 
	F.~S.~N.~Lobo,
	``Stability of phantom wormholes,''
	{\it Phys.\ Rev.\ D} {\bf 71} (2005) 124022
	doi:10.1103/PhysRevD.71.124022
	\href{https://arxiv.org/abs/gr-qc/0506001}{arXiv:gr-qc/0506001}
	
	\bibitem{Eiroa:2007qz} 
	E.~F.~Eiroa and C.~Simeone,
	``Stability of Chaplygin gas thin-shell wormholes,''
	{\it Phys.\ Rev.\ D} {\bf 76}, 024021 (2007)
	\href{https://arxiv.org/abs/0704.1136}{arXiv:0704.1136 [gr-qc]}
	
	\bibitem{Gorini:2008zj}
	V.~Gorini, U.~Moschella, A.~Y.~Kamenshchik, V.~Pasquier and A.~A.~Starobinsky,
	``Tolman-Oppenheimer-Volkoff equations in presence of the Chaplygin gas: stars and wormhole-like solutions,''
	{\it Phys.\ Rev.\ D} {\bf 78} (2008) 064064
	\href{https://arxiv.org/abs/0807.2740}{arXiv:0807.2740 [astro-ph]}
	
	\bibitem{Halilsoy:2013iza} 
	M.~Halilsoy, A.~Ovgun and S.~H.~Mazharimousavi,
	``Thin-shell wormholes from the regular Hayward black hole,''
	{\it Eur.\ Phys.\ J.\ C }{\bf 74} (2014) 2796
	\href{https://arxiv.org/abs/1312.6665}{arXiv:1312.6665 [gr-qc]}
	
	\bibitem{Harko:2014oua} 
	T.~Harko, F.~S.~N.~Lobo and M.~K.~Mak,
	``Wormhole geometries supported by quark matter at ultra-high densities,''
	{\it Int.\ J.\ Mod.\ Phys.\ D} {\bf 24} (2014) no.01,  1550006
	\href{https://arxiv.org/abs/1403.0771}{arXiv:1403.0771 [gr-qc]}
		
    \bibitem{Garriga:1991ts}
    J.~Garriga and A.~Vilenkin,
    ``Perturbations on domain walls and strings: A Covariant theory,''
    {\it Phys.\ Rev.\ D} {\bf 44} (1991) 1007.
    doi:10.1103/PhysRevD.44.1007
    
    \bibitem{Brax:2001qd} 
    P.~Brax, C.~van de Bruck and A.~C.~Davis,
    ``Brane world cosmology, bulk scalars and perturbations,''
    {\it JHEP}{\bf 0110} (2001) 026
    \href{https://arxiv.org/abs/hep-th/0108215}{arXiv: hep-th/0108215}
    
    \bibitem{Guven:1993ew}  
    J.~Guven,
    ``Covariant perturbations of domain walls in curved space-time,''
    {\it Phys.\ Rev.\ D} {\bf 48} (1993) 4604
    \href{https://arxiv.org/abs/gr-qc/9304032}{arXiv:gr-qc/9304032}
    
    \bibitem{Ishibashi:2002nn} 
    A.~Ishibashi and T.~Tanaka,
    ``Can a brane fluctuate freely?,''
    {\it JCAP} {\bf 0503} (2005) 011
    \href{https://arxiv.org/abs/gr-qc/0208006}{arXiv:gr-qc/0208006}
    
    \bibitem{Boehm:2002kf} 
    T.~Boehm and D.~A.~Steer,
    ``Perturbations on a moving D3-brane and mirage cosmology,''
    {\it Phys.\ Rev.\ D} {\bf 66}, 063510 (2002)
    \href{https://arxiv.org/abs/hep-th/0206147}{arXiv:hep-th/0206147}
    
    \bibitem{Deruelle:2000yj}  
    N.~Deruelle, T.~Dolezel and J.~Katz,
    ``Perturbations of brane worlds,''
    {\it Phys.\ Rev.\ D} {\bf 63} (2001) 083513
    \href{https://arxiv.org/abs/hep-th/0010215}{arXiv:hep-th/0010215}
    
    \bibitem{Brax:2002nx} 
    P.~Brax, D.~Langlois and M.~Rodriguez-Martinez,
    ``Fluctuating brane in a dilatonic bulk,''
    {\it Phys.\ Rev.\ D} {\bf 67} (2003) 104022
    \href{https://arxiv.org/abs/hep-th/0212067}{arXiv:hep-th/0212067}
    
    \bibitem{Hayward:2005gi}
    S.~A.~Hayward,
    ``Formation and evaporation of regular black holes,''
    {\it Phys.\ Rev.\ Lett.\ } {\bf 96} (2006) 031103
    \href{https://arxiv.org/abs/gr-qc/0506126}{arXiv:gr-qc/0506126}
    
    \bibitem{Frolov:2017dwy}
    V.~P.~Frolov,
    ``Remarks on non-singular black holes,''
    {\it EPJ Web Conf.\ } {\bf 168} (2018) 01001
    \href{https://arxiv.org/abs/1708.04698}{arXiv:1708.04698 [gr-qc]}
    
    \bibitem{Frolov:2017rjz}
    V.~P.~Frolov and A.~Zelnikov,
    ``Quantum radiation from an evaporating nonsingular black hole,''
    {\it Phys.\ Rev.\ D} {\bf 95} (2017) no.12,  124028
    \href{https://arxiv.org/abs/1704.03043}{arXiv:1704.03043 [hep-th]}
   
   	\bibitem{Israel:1966rt}
   W.~Israel,
   ``Singular hypersurfaces and thin shells in general relativity,''
   {\it Nuovo Cim.\ B} {\bf 44S10} (1966) 1
   {\it Nuovo Cim.\ B } {\bf 44} (1966) 1 
   {\it Erratum: Nuovo Cim.\ B} {\bf 48} (1967) 463
   doi:10.1007/BF02710419, 10.1007/BF02712210
   
   \bibitem{Gibbons:1976ue}
   G.~W.~Gibbons and S.~W.~Hawking,
   ``Action Integrals and Partition Functions in Quantum Gravity,''
   {\it Phys.\ Rev.\ D} {\bf 15} (1977) 2752.
   doi:10.1103/PhysRevD.15.2752
   
   \bibitem{Chamblin:1999ya} 
   H.~A.~Chamblin and H.~S.~Reall,
   ``Dynamic dilatonic domain walls,''
   {\it Nucl.\ Phys.\ B} {\bf 562}, 133 (1999)
   \href{https://arxiv.org/abs/hep-th/9903225}{arXiv:9903225 [hep-th]}	
   
\end{thebibliography}
\end{document}